\documentclass[english,aps,prX,showpacs,twocolumn]{revtex4-2}
\usepackage[T1]{fontenc}
\usepackage[utf8]{inputenc}
\usepackage{gensymb}
\usepackage[english]{babel}
\usepackage{graphicx}
\usepackage{bm}
\usepackage{bbm}
\usepackage{array}
\usepackage{amsmath}
\usepackage{amssymb}
\usepackage{dcolumn}
\usepackage{epstopdf}
\usepackage{bbold}
\usepackage{tikz}
\usepackage{pgfplots}
\usepackage[margin=15mm]{geometry}
\colorlet{linkequation}{blue}
\usepackage[colorlinks=true,linkcolor=blue,citecolor=blue,urlcolor=blue]{hyperref}
\usetikzlibrary{shadings}
\usepackage[nohyperlinks, nolist]{acronym}
\usepackage{braket}
\usepackage{mathrsfs}
\usepackage{import}
\usepackage{filecontents}
\usepackage{color}

\renewcommand{\vec}[1]{\boldsymbol{#1}}

\newcommand{\unitvec}[1]{\hat{\vec{\mathrm{#1}}}}
\newcommand{\er}{\unitvec e_r}
\newcommand{\parent}[1]{\bigg(#1\bigg)}
\newcommand{\etheta}{\unitvec e_\vartheta}
\newcommand{\ephi}{\unitvec e_\varphi}

\begin{document}

\title{Nutational switching in ferromagnets and antiferromagnets
}

\author{Lucas Winter}

\affiliation{Fachbereich Physik, Universit\"at Konstanz, DE-78457 Konstanz, Germany}

\author{Sebastian Gro{\ss}enbach}
\affiliation{Fachbereich Physik, Universit\"at Konstanz, DE-78457 Konstanz, Germany}

\author{Ulrich Nowak}
\affiliation{Fachbereich Physik, Universit\"at Konstanz, DE-78457 Konstanz, Germany}

\author{Levente R{\'o}zsa}
\affiliation{Fachbereich Physik, Universit\"at Konstanz, DE-78457 Konstanz, Germany}
\email[]{levente.rozsa@uni-konstanz.de}

\date{\today}

\renewcommand{\figurename}{FIG.}


\begin{abstract}
    It was demonstrated recently that on ultrashort time scales magnetization dynamics does not only exhibit precession but also nutation. Here, we investigate how nutation can contribute to spin switching leading towards ultrafast data writing. 
    We use analytic theory and atomistic spin
simulations to discuss the behavior of ferromagnets and antiferromagnets in high-frequency magnetic fields. In ferromagnets, linearly polarized fields align the magnetization perpendicular to the external
field, enabling 90\degree~switching. For circularly polarized fields in the $xy$
plane, the magnetization tilts to the $z$ direction. During this tilting,
it rotates around the $z$ axis, allowing 180\degree~switching. In antiferromagnets, external fields with frequencies higher than the nutation frequency align the order parameter parallel to the field direction, while
for lower frequencies it is oriented perpendicular to the field.
The switching frequency increases with the magnetic field
strength, and it deviates from the Larmor frequency, making it possible to outpace precessional switching in high magnetic fields.
\end{abstract}

\maketitle

\begin{acronym}
\acro{DMI}[DMI]{Dzyaloshinsky-Moriya interaction}
\acro{sLLG}[sLLG]{stochastic Landau-Lifshitz-Gilbert}
\acro{LLG}[LLG]{Landau-Lifshitz-Gilbert}
\end{acronym}

\section{Introduction}

Deterministic spin switching plays an important role in industrial applications, for example in magnetic memory devices~\cite{stanciu_all-optical_2007,vahaplar_ultrafast_2009,radu_transient_2011,vahaplar_all-optical_2012}. However, as industrial needs exceed currently available switching speeds, the a search for new physical effects that can accelerate spin switching is ongoing. One such pathway might be exploiting spin nutation.  
In spin nutation, the direction of the magnetic moment and angular momentum become separated on ultrashort time scales~\cite{ciornei_magnetization_2011, wegrowe_magnetization_2012}, and the magnetic moment rotates around the angular momentum~\cite{bottcher_significance_2012}. Spin nutation is described by the inertial Landau-Lifshitz-Gilbert (ILLG) equation~\cite{ciornei_magnetization_2011, wegrowe_magnetization_2012}. The ILLG equation extends the Landau-Lifshitz-Gilbert (LLG) equation by an additional inertial term containing a second-order time derivative of the magnetic moment. The inertial term originates from spin-orbit coupling as it can be derived using the Dirac equation~\cite{mondal_relativistic_2017}.

Linear-response theory applied to the ILLG equation predicts a spin nutation resonance in addition to the precession resonance in ferromagnets~\cite{olive_deviation_2015, olive_beyond_2012,cherkasskii_nutation_2020} and in antiferromagnets~\cite{mondal_nutation_2021}. Recently the nutation resonance has been experimentally observed in ferromagnetic thin films~\cite{neeraj_inertial_2021}. Estimates for the angular momentum relaxation parameter $\eta$ characterizing the nutation period range from the order of 1~fs~\cite{ciornei_magnetization_2011,bhattacharjee_atomistic_2012,li_inertial_2015,thonig_magnetic_2017} to several 100~fs~\cite{neeraj_inertial_2021, unikandanunni_inertial_2021}. Experiments in epitaxial cobalt films suggest that this parameter is proportional to the magnetocrystalline anisotropy~\cite{unikandanunni_inertial_2021}, which similarly emerges due to the spin-orbit coupling.

For writing data in magnetic hard drives, usually a magnetic pulse with opposite direction to the magnetization is applied~\cite{Mallinson_Damped, Bertotti_Comparison}. Subsequently, the damping switches the magnetization in approximately 100 ps. In this case, the damping constant limits the switching time~\cite{Gilbert_IEEE}. Several authors have proposed faster switching mechanisms by using transverse magnetic field pulses~\cite{gerrits_ultrafast_2002,kaka_precessional_nodate}. In this case, the switching time depends on the energy delivered by the magnetic pulse. Tudosa \textit{et al.} experimentally demonstrated ultrafast switching using magnetic field pulses~\cite{tudosa_ultimate_2004}. Moreover, they argued that switching on a time scale shorter than 2 ps is impossible in ferromagnets. However, because of practical limitations, this lower bound has never been confirmed experimentally ~\cite{tudosa_ultimate_2004}. By using the intrinsic inertia of antiferromagnetic dynamics, it is possible to reduce switching times by a factor of ten compared to ferromagnets~\cite{kimel_inertia-driven_2009}. The role of the extrinsic inertia as described by the ILLG equation for spin switching has hardly been studied so far, apart from a recent work by Neeraj \textit{et al.}~\cite{neeraj_magnetization_2021} which focused on applying a magnetic field pulse without an oscillating component. They showed that the inertial term
opens an additional energy channel slightly modifying the switching time. However, it remains to be seen how a resonant excitation of the nutational motion influences the switching time. Since the nutation frequency is much higher than the precession frequency, it stands to reason that nutation could be advantageous for faster magnetization switching.

Here we investigate how terahertz magnetic fields can be utilized for spin switching in ferromagnets (FMs) and antiferromagnets (AFMs). We show that if the nutation is resonantly excited by the magnetic field, it exerts a torque on the order parameter, reorienting its direction. We discuss nutational switching modes analytically by proposing an alternative form of the ILLG equation based on the angular momentum, and numerically by using atomistic spin simulations. It is found that a linearly polarized magnetic field enables a $90\degree$ switching mode of the order parameter. A circularly polarized field initiates the tilting of the order parameter from the equilibrium direction earlier than the linearly polarized field, and it can also be used for $180\degree$ spin switching. The switching time is found to be around 20~ps in FMs and around 5~ps for AFMs. It is demonstrated that the switching velocity of nutational switching is by a factor of two higher than that of precessional switching in a wide parameter range.

The paper is organized as follows. In Sec.~\ref{sec:analyticTreatment}, we treat nutational switching analytically, and propose an alternative representation of the ILLG equation using the angular momentum. In this representation, we discuss nutational switching in FMs in Sec.~\ref{sec:Ferromagnets} both by linearly and by circularly polarized fields. Using atomistic spin simulations, we show that linearly polarized fields can be used for $90\degree$ switching and circularly polarized fields for $180\degree$ switching. We investigate nutational switching in AFMs in Sec.~\ref{sec:Antiferromagnets}. We determine the switching time for a wide range of material parameters and compare it to the switching time in FMs and to precessional switching. 

\section{Theory of nutational switching} \label{sec:analyticTreatment}

The dynamics of the magnetic moments will be described using the ILLG equation as derived in earlier works~\cite{ciornei_magnetization_2011,bhattacharjee_atomistic_2012,mondal_relativistic_2017}, reading
\begin{align}
\begin{split}
       \vec{\dot{M}}_i =& -\gamma_i \vec M_i \times \vec H_{\mathrm{eff},i} + \frac{\alpha_i}{M_{0, i}}\vec M_i \times \vec{\dot{M}}_i \\ &+ \frac{\eta_i}{ M_{0, i}} \vec M_i \times \vec{\ddot{M}}_i. \label{eq:ILLG}
\end{split}
\end{align}

The first term in Eq.~\eqref{eq:ILLG} describes precession with the gyromagnetic ratio $\gamma_i$ and the second term  transversal relaxation with damping constant $\alpha_{i}$. The third term induces inertial dynamics with inertial relaxation time $\eta_i$. Here, $i$ enumerates the magnetic moments, taking into account possible deviations in the gyromagnetic ratio and the relaxation time between different sublattices. $M_{0, i}$ is the magnitude of the magnetic moment and $\vec H_{\mathrm{eff},i}$ denotes the effective field $\vec H_{\mathrm{eff},i} = -\frac{\partial \mathcal H }{\partial \vec M_i}$, where $\mathcal H=\mathcal{H}_{\textrm{exch}}+\mathcal{H}_{\textrm{ani}}-\sum_{i}\vec{M}_{i}\left(\vec{H}_{\textrm{ext}}+\vec{H}_{\textrm{osc}}(t)\right)$ is the Hamiltonian of the system containing exchange $\mathcal{H}_{\textrm{exch}}$ and anisotropy $\mathcal{H}_{\textrm{ani}}$ energy as well as static $\vec{H}_{\textrm{ext}}$ and oscillating $\vec{H}_{\textrm{osc}}$ external field terms. The $\mathcal{H}_{\textrm{exch}}$ and $\mathcal{H}_{\textrm{ani}}$ terms will be specified in Secs.~\ref{sec:Ferromagnets} and \ref{sec:Antiferromagnets} for FMs and AFMs, respectively. In the simulations, we solve the ILLG equation numerically using the implementation of Heun's method described in Ref.~\cite{mondal_nutation_2021}.

\begin{figure}
    \centering
    \includegraphics[width = 0.95\columnwidth]{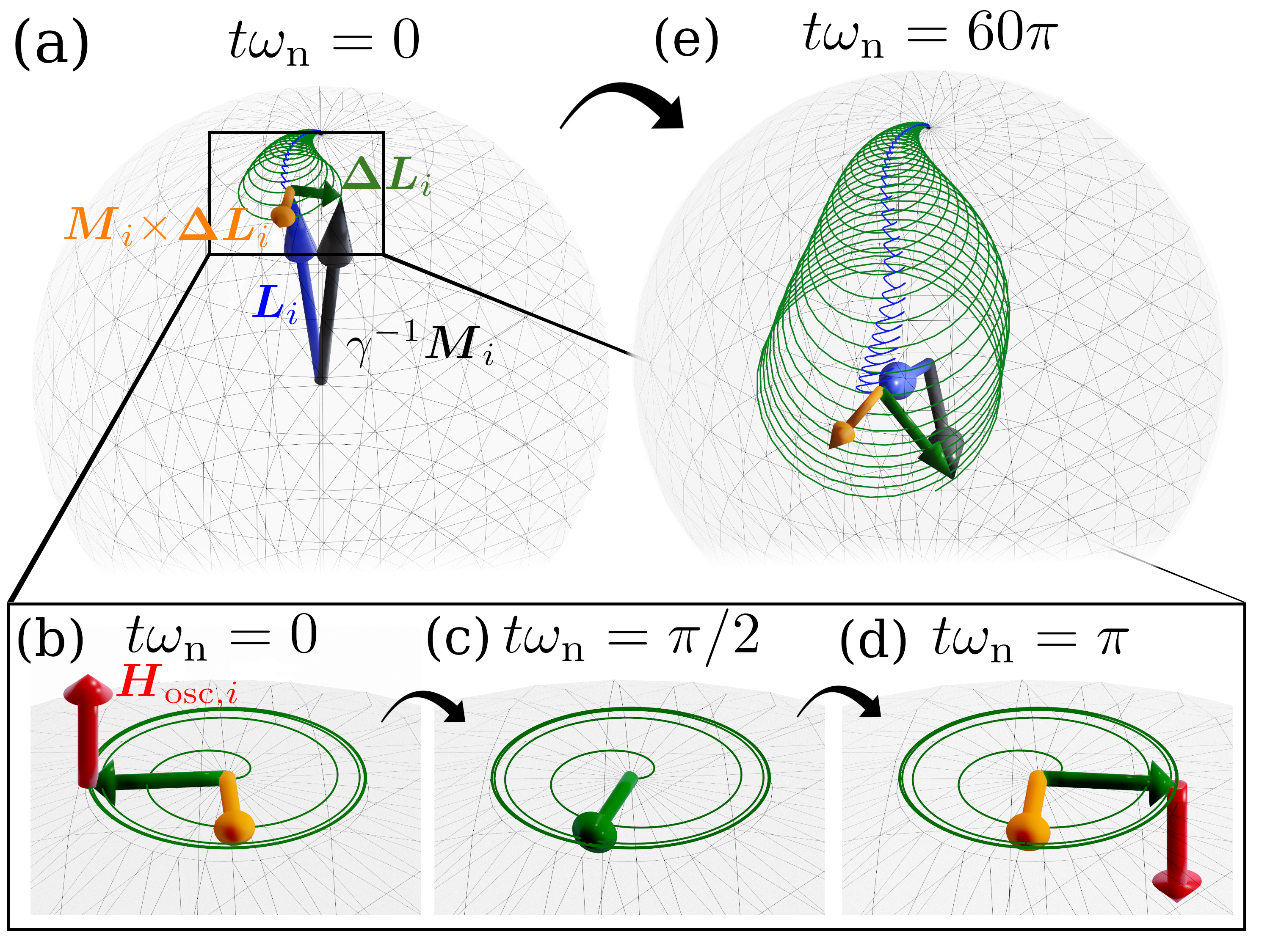}
    \caption{Illustration of nutational switching. (a) First, an oscillating magnetic field $\vec H_{\textrm{osc}}$ is applied to a magnetic moment $\vec M_{i}$. (b)-(d) This excites a finite nutation amplitude $\Delta \vec L_{i}$. (e) The nutation vector in conjunction with the external field exerts a torque proportional to $\langle \Delta \vec L_{i} \times \vec H_{\textrm{osc}}\rangle_t$. 
}
    \label{fig:nutswitchexplain}
\end{figure}

Because of the inertial term, the angular momentum $\vec L_i$ is no longer parallel to the magnetic moment $\vec{M}_{i}$ as in the LLG equation~\cite{ciornei_magnetization_2011}, leading to
\begin{align}
    \vec L_i  = \frac{1}{\gamma_i}\vec M_i - \Delta \vec L_i. \label{eq:angularMomentum}
\end{align}
Here we introduced the nutation vector $\Delta \vec L_i = \frac{\eta_i}{M_{0,i}\gamma_i} \vec M_i \times \vec{\dot{M}}_i$ for this deviation. The vectors $\vec L_i, \vec M_i$ and $\Delta \vec L_i$ are illustrated in Fig.~\ref{fig:nutswitchexplain}(a). Note that for simplicity we used a sign convention where without the inertial term $\vec{L}_i$ is pointing along $\vec{M}_i$ ($\gamma_{i}>0$), although Eq.~(\ref{eq:ILLG}) describes the time evolution of electronic spins. 
The nutation amplitude $|\Delta \vec L_{i}|$ is usually much smaller than the angular momentum amplitude, because $|\Delta \vec L_{i}| = |\frac{\eta}{\gamma M_{0,i}} \vec M_{i} \times \dot{\vec M}_{i} | \approx \frac{\eta}{M_{0,i}}| \vec M_{i} \times \left(\vec M_{i} \times \vec H_{\mathrm{eff},i}\right)| \leq \gamma_{i} \eta_{i} |\vec L_{i}|$ and $\gamma_{i} \eta_{i} \ll 1$ even for relatively large inertial relaxation times of $\eta_{i}\approx 100$~fs. 

For understanding the nutation-induced switching, we decompose the magnetic moment using the angular momentum $\vec L_i$ and the nutation $\Delta \vec L_i$ according to $\vec M_i = \gamma(\vec L_i + \Delta \vec L_i)$ . The ILLG equation becomes a system of 
coupled first-order differential equations,
\begin{align}
    \vec{\dot{L}}_i =& - \gamma_{i} \vec L_{i} \times \vec H_{\mathrm{eff},i} - \gamma_{i} \Delta \vec L_i \times \vec H_{\mathrm{eff}, i} + \frac{\alpha_i}{ \eta_i} \Delta \vec L_i, \label{eq:dldt} \\
    \begin{split}
    \Delta \vec{\dot{L}}_i =& \gamma_i \Delta \vec L_i \times \left(\frac{1}{M_{0,i} \eta_i}\vec L_i  + \vec H_{ \mathrm{eff}, i} \right) \\ &+\gamma_{i} \vec L_{i} \times \vec H_{i, \mathrm{eff}} - \frac{\alpha_i}{\eta_i} \Delta \vec L_i; \label{eq:ddeltaLdt}\end{split}
\end{align}
see Appendix~\ref{subsec:ILLGAngular} for the derivation.

We will analyze these differential equations in the following, starting with a qualitative understanding of nutational switching. 
A field $\vec H_{\mathrm{osc}}$ oscillating at the nutation frequency $\omega = \omega_{\mathrm n}$ is applied to the magnetic moment, as shown in Fig.~\ref{fig:nutswitchexplain}(a). As a consequence, the magnetic moment starts to nutate with finite nutation amplitude. Linear-response theory predicts nutation with the same frequency as $\vec H_{\mathrm{osc}}$ and a finite phase shift. However, according to Eq.~\eqref{eq:dldt}, $\Delta \vec L_{i}$ will interact with the oscillating field $\vec H_{\mathrm{osc}}$ and exert a torque $-\gamma \Delta \vec L_{i} \times \vec H_{\mathrm{osc}}$ on the angular momentum $\vec L_{i}$. Figure~\ref{fig:nutswitchexplain}(b) shows the direction of the torque. Because of the constant phase difference between $\Delta \vec L_{i}$ and $\vec H_{\mathrm{osc}}$, the torque still points in the same direction after half a nutation period; see Fig.~\ref{fig:nutswitchexplain}(d). This means that the torque has a finite time average that drives the switching, as shown in Fig.~\ref{fig:nutswitchexplain}(e). 

Theoretical estimates~\cite{ciornei_magnetization_2011,bhattacharjee_atomistic_2012,li_inertial_2015,thonig_magnetic_2017} supported by recent experimental evidence~\cite{neeraj_inertial_2021, unikandanunni_inertial_2021} suggest that the nutation frequency $\eta_{i}^{-1}$ is much higher than the precession frequency. This allows for a separation of time scales between the fast variable $\Delta \vec{L}_{i}$ and the slow variable $\vec{L}_{i}$. We define a coordinate system where the angular momentum is oriented along the radial direction in spherical coordinates $\vec L_i = L_{0, i} \cdot \unitvec{e}_{r, i}$. The nutation is then essentially a two-dimensional motion happening in the plane perpendicular to the angular momentum as the nutation amplitude is usually small. We introduce the complex amplitude $c_i(t)$ for the nutation vector $\Delta \vec L_i$ with $\Delta \vec L_i =  \mathrm{Re}\{c_i(t)\} \unitvec{e}_{\vartheta, i} + \mathrm{Im}\{c_i(t)\} \unitvec{e}_{\varphi, i}$ in the coordinate system defined by the angular momentum. 
This allows for finding an instantaneous solution for the nutation amplitude $c_i(t)$ in a linear-response framework; see Appendix \ref{sec:difEquationNutation}.

For a single macrospin, we can give the closed-form solution for $\vec H_{\mathrm{eff}}(t, \vartheta, \varphi) = \vec H_{0}(\vartheta,\varphi) + \vec H_{\mathrm{osc}}(t,\vartheta, \varphi)$, a sum of a time-dependent oscillating field $\vec H_{\mathrm{osc}} = \vec h e^{-i\omega t} + \vec h^* e^{i\omega t}$ with complex amplitude $\vec h$ and a time-independent external field $\vec H_{0}(\vartheta, \varphi)$ which contains the external Zeeman field $\vec{H}_{\textrm{ext}}$ and anisotropy $\vec{H}_{\textrm{ani}}$. The contribution of $\vec H_{\mathrm{osc}}$ to $\omega_{\textrm{n}}$ is neglected in the linear-response regime. From now on we will not explicitly write the orientation dependence $\left(\vartheta,\varphi\right)$. With these notations one obtains
\begin{align}
\begin{split}
    c(t) = &c_{\mathrm{hom}}(t) - M_0(\unitvec e_\varphi - i\etheta)\cdot \bigg(\frac{e^{-i\omega t}}{i(\omega_{\mathrm{n}} - \omega) + \alpha/\eta}\ \vec h \\ &+ \frac{e^{i\omega t}}{i(\omega_{\mathrm{n}} + \omega) + \alpha/\eta}\ \vec h^{*} + \frac{1}{i\omega_{\mathrm{n}} + \alpha/\eta }\vec H_0 \bigg), \label{eq:nutationClosedForm}
\end{split}
\end{align}
with the homogeneous solution $c_{\mathrm{hom}} = c_0 e^{-\frac i \eta t} e^{-\frac \alpha \eta t}$. For $t \gg \eta/\alpha$, the homogeneous solution will become negligible. In this case, the nutation is a circular motion with the frequency $\omega$ of the external field and a resonance at $\omega=\omega_{\textrm{n}}$. Moreover, the anisotropy and the constant magnetic field are responsible for a constant offset of the nutation amplitude. This offset leads to the relaxation of the magnetic moment towards the direction of the angular momentum. \\

Using Eq.~\eqref{eq:ddeltaLdt}, it is now possible to discuss the time evolution of the angular momentum $\vec L$ taking place on longer time scales.
We can express the change in angular momentum \eqref{eq:dldt} using the nutation vector $\Delta \vec L$ as
\begin{align}
    \langle \vec{\dot{L}} \rangle_t = -\langle \gamma \vec L \times \vec H_{\mathrm{eff}} \rangle_t+\left< \frac{\alpha}{\eta} \Delta \vec L \right>_t  -\langle \gamma \Delta \vec L \times \vec  H_{\mathrm{eff}} \rangle_t, \label{eq:dldtaverage} 
\end{align}
where $\braket{\cdot }_t$ indicates time averaging over one nutation period. The first term in Eq.~\eqref{eq:dldtaverage} is the precession term, which is functionally unchanged from the LLG equation. The second term in Eq.~\eqref{eq:dldtaverage} is responsible for the damping. In this representation, the damping is closely linked with the nutation amplitude. For a more detailed explanation see Appendix \ref{sec:nutationDamping}.
The third term in Eq.~\eqref{eq:dldtaverage}, $-\langle \gamma \Delta \vec L \times \vec  H_{\mathrm{eff}} \rangle_t$, only appears because of inertial effects. This term is responsible for the nutational switching.

\section{Nutational switching in ferromagnets} \label{sec:Ferromagnets}

\subsection{Nutational switching for linearly polarized fields}

\begin{figure}
    \centering
    \includegraphics[ width =\columnwidth]{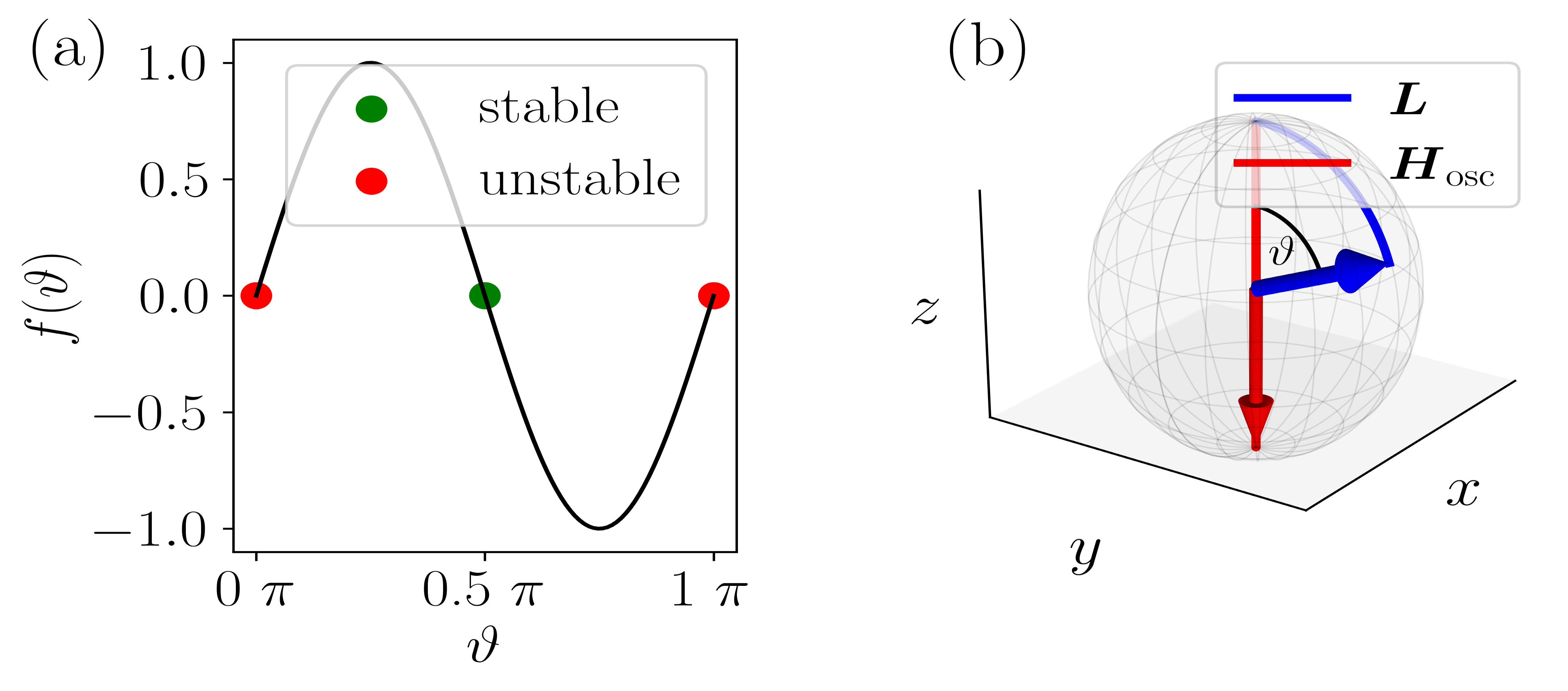}
    \caption{
(a) Dependence of the right-hand side of Eq.~\eqref{eq:tilting} on $\vartheta$. The stationary point $\vartheta = \pi/2$ is stable, $\vartheta = 0, \pi$ are unstable. 
(b) Illustration of the solution of the differential equation~\eqref{eq:tilting}. The linearly polarized field along the $z$-axis tilts the spin towards the $y$-direction. 
}
    \label{ana:SingleLinear}
\end{figure}
For an easier interpretation, we will neglect the precession and damping terms in Eq.~\eqref{eq:dldtaverage} in the following. We describe the time evolution of $\vec L$ on this longer time scale in spherical coordinates, where $\vartheta$ represents the angle between $\vec L$ and the $z$-axis and $\varphi$ represents the angle measured from the $x$-axis in the $xy$-plane.

For a linearly polarized field $\vec H_{\mathrm{osc}} = h \unitvec e_z \cos(\omega t)$, Eq.~\eqref{eq:dldtaverage} can be expressed as 
\begin{align}
    \dot \vartheta = \frac{1}{2}\parent{\frac{\gamma h}{2}}^2  \frac{\alpha/\eta}{(\omega_{\mathrm{n}} - \omega)^2 + \frac{\alpha^2}{\eta^2} }\sin(2\vartheta). \label{eq:tilting}
\end{align}

For an illustration of the right-hand side of Eq.~\eqref{eq:tilting}, see Fig.~\ref{ana:SingleLinear}(a). There are three stationary points. $\vartheta = 0$ and $\vartheta = \pi$ are unstable. In these cases, the spin is either parallel or antiparallel to the magnetic field, but any small perturbation will tilt it away from this direction. The point $\vartheta = \pi/2$ is stable. In this case, the spin is perpendicular to the magnetic field. If the spin is not in one of these three configurations, it will align itself perpendicular to the field at $\vartheta = \pi/2$, as shown in Fig.~\ref{ana:SingleLinear}(b). This type of switching is shown in Supplemental Video 1~\cite{supp}.

From Eq.~\eqref{eq:tilting} it follows that the tilting velocity is highest at the nutation resonance $\omega = \omega_{\mathrm{n}}$. Materials with high values of $\eta$ and low values of $\alpha$ should exhibit the highest tilting speeds. Moreover, the tilting velocity increases with increasing magnetic field, and the magnetic moment will align perpendicular to a linearly polarized magnetic field. In a system with cubic anisotropy $\mathcal H_{\textrm{ani}} = -K_{4}(M_x^4 + M_y^4 + M_z^4)/M_0^4$ with potential minima separated by an angle of $90\degree$, applying the field along the magnetization direction should enable $90\degree$ switching. Such a switching would be possible to detect by measuring the magnetization direction or through the anisotropic magnetoresistance. 

\begin{figure}
    \centering
    \includegraphics[width=\columnwidth]{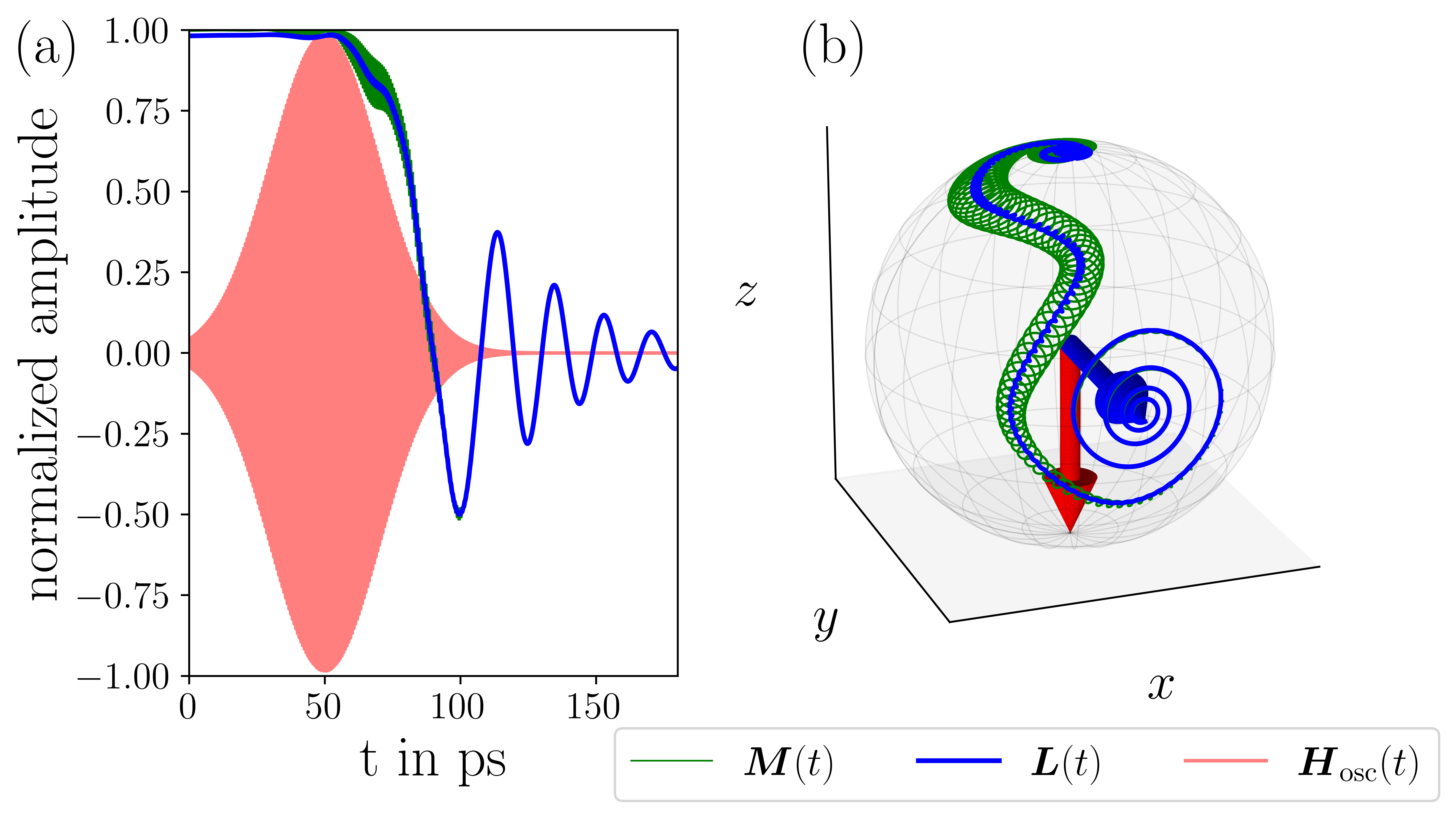}
    \caption{$90\degree$ switching using a linearly polarized field along the $z$-direction. The material parameters are $\alpha=0.1, \eta =100$~fs and a cubic anisotropy with $K_{4}=10^{-23}\ $J. The magnetic field strength is $h=5$~T, the field frequency is $\omega=\omega_{\mathrm{n}}=1\cdot 10^{13}$~s$^{-1}$ and the pulse width is $\mathrm{FWHM} = 47$~ps. 
}
    \label{fig:ferroLinearSwitch}
\end{figure}

We use atomistic spin simulations to validate this prediction and start with a single macrospin aligned close to the $z$-direction. We then apply a magnetic field pulse along the $z$-direction at the nutation frequency $\omega = \omega_{\mathrm{n}} \approx \frac 1 \eta$ with a Gaussian pulse shape given by
$\vec H_{\mathrm{osc}} = h \exp\left(-\frac{(t - t_0)^2}{2\sigma^2}\right)\unitvec e_z \cos(\omega t)$ with $\sigma = 2\sqrt{2\ln(2)}\mathrm{FWHM}$.
As shown in Fig.~\ref{fig:ferroLinearSwitch}, 
the behavior of the magnetic moment in the simulations is consistent with the analytic prediction shown in Fig.~\ref{ana:SingleLinear}, but superimposed with a precession caused by the anisotropy term. Here, the switching was achieved with a pulse width of $\mathrm{FWHM} = 47$~ps. The tilting speed itself is high as it takes only 23~ps for $L_z$ to decay to $1/\textrm{e}$ of its initial value. If the magnetic field is turned off at this point, one can be sure that the magnetic moment will relax to the switched position.
However, it takes a significant amount of time for the tilting to start, because a magnetic field almost parallel to the magnetic moment excites the nutation slowly according to Eq.~\eqref{eq:tilting}. Applying the linearly polarized field at a small angle from the initial equilibrium direction could likely decrease the switching time. Moreover, finite temperature would likely alleviate this problem because there the spin would not be exactly parallel to the external field. 

\subsection{Nutational switching for circularly polarized fields}

\begin{figure}
    \centering
    \includegraphics[ width = \columnwidth]{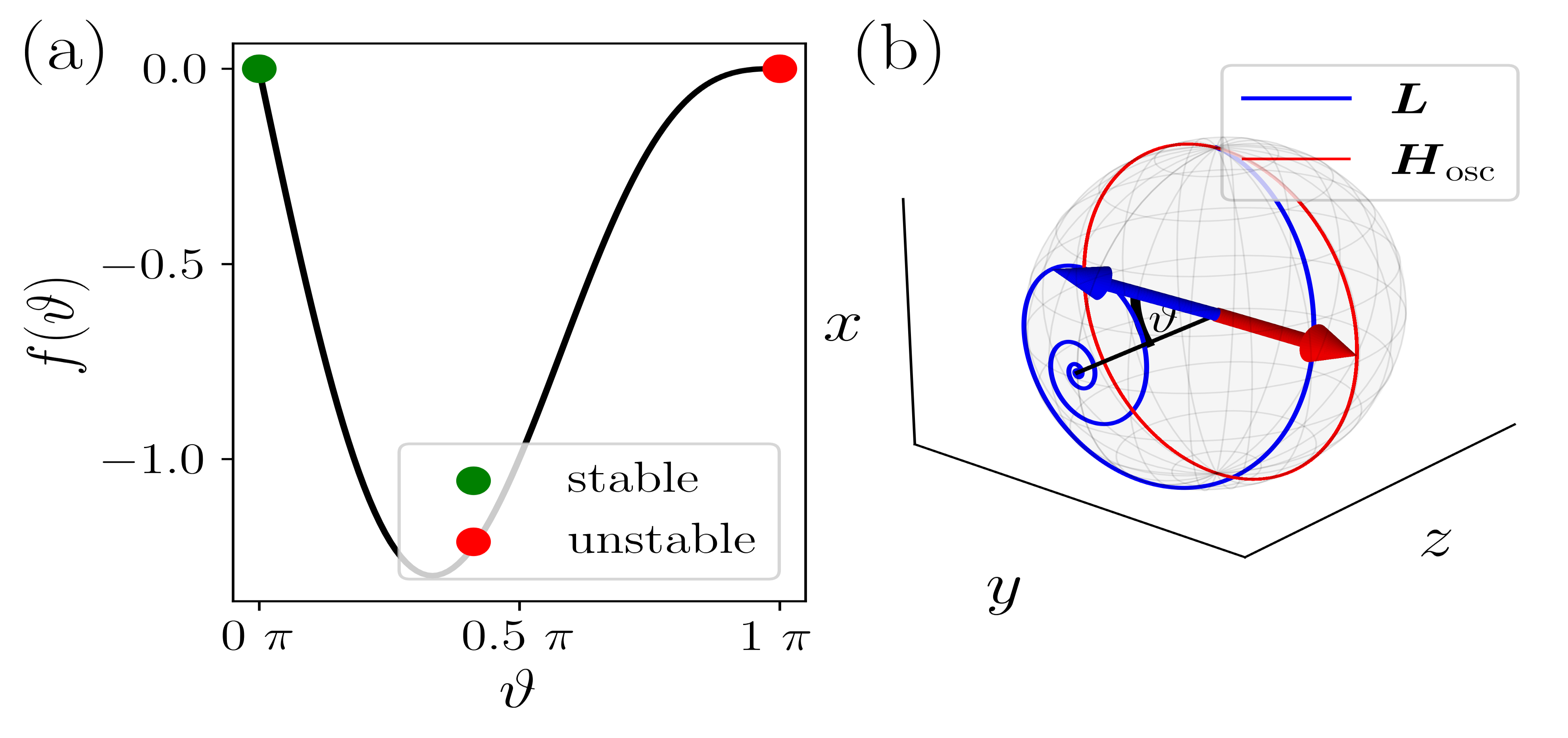}
    \caption{ 
(a) Dependence of the right-hand side of Eq.~\eqref{de:ThetaCircular} on $\vartheta$. The stationary point $\vartheta = 0$ is stable, $\vartheta = \pi$ is unstable.
(b) Illustration of the solution to the differential equations~\eqref{de:ThetaCircular} and \eqref{de:PhiCircular}. In the beginning, the magnetic moment points along the $x$-direction.
}
    \label{ana:SingleCircular}
\end{figure}

For a circularly polarized field $\vec H_{\mathrm{osc}} = h \left[ \cos(\omega t)\ \unitvec e_x +  \sin(\omega t)\ \unitvec e_y \right]$, Eq.~\eqref{eq:dldtaverage} can be expressed in spherical coordinates as
 \begin{align}
     \dot \vartheta &= -\frac{(\gamma h)^2}{4}\frac{\alpha/\eta}{(\omega_{\mathrm{n}} - \omega)^2 + \frac{\alpha^2}{\eta^2}}\left(\sin(\vartheta) + \frac{1}{2}\sin(2\vartheta)\right) \label{de:ThetaCircular}, \\
     \dot \varphi &= -\frac{(\gamma h)^2}{2}  \frac{\omega_{\mathrm{n}} - \omega}{(\omega_{\mathrm{n}} - \omega)^2 + \frac{\alpha^2}{\eta^2}} \big(1+\cos(\vartheta)\big).\label{de:PhiCircular}
 \end{align}

Equation~\eqref{de:ThetaCircular} decouples from Eq.~\eqref{de:PhiCircular}. Equation~\eqref{de:PhiCircular} shows that the angular momentum rotates around the $z$-axis. The sense of rotation depends on whether the frequency of the external field is higher or lower than the nutation frequency. In resonance, $\omega = \omega_{\mathrm{n}}$, the rotation around the $z$-axis vanishes. The maximal rotation frequency is achieved for $\omega = \omega_{\mathrm{n}} \pm \alpha/\eta$. 

Figure~\ref{ana:SingleCircular}(a) shows the right-side of Eq.~\eqref{de:ThetaCircular}. There are two stationary points. First, we can see that $\vartheta = 0$ is stable, as the derivative on the right-hand side is negative. The point $\vartheta = \pi$ is unstable against perturbations. For all other angles, the magnetic moment rotates to align itself parallel to the $z$-direction. As for linearly polarized magnetic fields, the speed is proportional to $h^2$. In resonance the tilting velocity is maximal. This type of switching is illustrated in Fig.~\ref{ana:SingleCircular}(b) and Supplemental Video 2~\cite{supp}.

The rotation around the $z$-axis can also be used for $180\degree$ switching. In this case, it is useful to maximize the rotation frequency $\dot\varphi$ and minimize the tilting speed. Because $\dot \varphi/\dot \vartheta \propto \frac{\eta(\omega_{\mathrm{n}} -\omega)}{\alpha}$, it is beneficial to consider a system with low damping and an excitation frequency above the nutation frequency. Circularly polarized fields could also be used for 90$\degree$ switching, using the tilting mechanism similarly to linearly polarized fields. In this case, the ratio $\dot \varphi/\dot \vartheta$ should be minimized. 

We use atomistic spin simulations to validate the analytical results. We apply a circularly polarized field in the $xy$-plane to a single macrospin. At $t = 0$~s the spin is aligned with the $y$-axis, selected by a uniaxial anisotropy term $\mathcal H_{\mathrm{ani}} = -\frac{K_y}{M_0^{2}} M_y^2$, facilitating 180$\degree$ switching. We apply an external field frequency above the nutation frequency and a smaller $\alpha$ value than in the linearly polarized case to maximize the ratio $\dot\varphi/\dot\vartheta$. 

\begin{figure}
    \centering
    \includegraphics[width=\columnwidth]{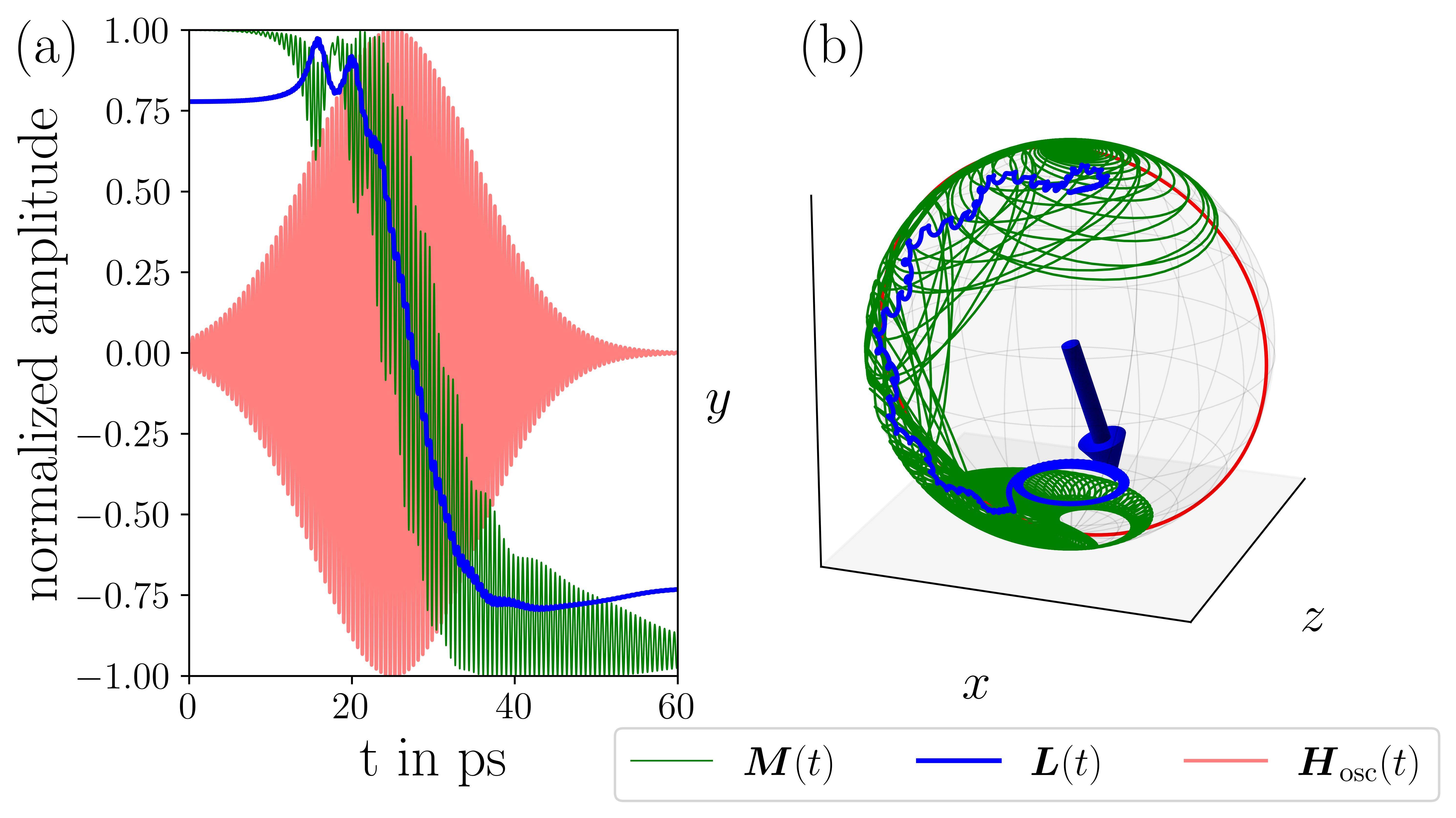}
    \caption{180$\degree$ switching using a circularly polarized field in the $xy$-plane. The material parameters are $\alpha=0.01, \eta =100$~fs, and a uniaxial anisotropy with $K_y = 10^{-23}$~J along the $y$-axis. The magnetic field strength is $h=5$~T, the field frequency is $\omega=1.1\cdot 10^{13}$~s$^{-1}$ and the pulse width is $\text{FWHM} = 23.75$~ps. 
 }
    \label{fig:ferroCircularSwitch}
\end{figure}

As shown in Fig.~\ref{fig:ferroCircularSwitch}, the behavior of the magnetization in the simulations is consistent with the analytical prediction shown in Fig.~\ref{ana:SingleCircular}(b). The nutation amplitude is much larger than in the linearly polarized case because $\alpha$ is smaller. Switching is possible with a pulse width of $\text{FWHM} = 23.75$~ps, significantly shorter than for linearly polarized fields. For a system with weaker anisotropy, even shorter pulses could be used. The switching starts significantly earlier than in the linearly polarized case because a nutation amplitude is excited much quicker. While a linearly polarized field parallel to the angular momentum does not lead to nutation, a circularly polarized field can excite a nutation in any orientation, according to Eq.~\eqref{eq:ddeltaLdt}.
 
\section{Nutational switching in antiferromagnets} \label{sec:Antiferromagnets}

\subsection{Analytical model}

In addition to the extrinsic inertial term $\eta_{i}$ of the ILLG equation, AFMs possess additionally an intrinsic inertia due to the exchange coupling between the two sublattices. This can be expected to have an impact on nutational switching. 

In AFMs the order parameter is the staggered magnetization $\vec N = \vec M_A - \vec M_B$. We can decompose this into a staggered nutation $\Delta \vec L_N = \Delta \vec L_A - \Delta \vec L_B$ and a staggered angular momentum $\vec L_N = \vec L_A - \vec L_B$ with $\vec N = \gamma(\vec L_N +  \Delta \vec L_N)$. As in FMs, the nutation vector of the sublattices can be represented by scalar complex variables $c_A$ and $c_B$ with $\Delta \vec L_A = \mathrm{Re}\{c_A\} \unitvec e_{\vartheta, A} + \mathrm{Im}\{c_A\} \unitvec e_{\varphi, A}$ and $\Delta \vec L_B = \mathrm{Re}\{c_B\} \unitvec e_{\vartheta, B} + \mathrm{Im}\{c_B\} \unitvec e_{\varphi, B}$. The Hamiltonian contains the exchange term $\mathcal{H}_{\textrm{exch}}=\frac{J}{M_0^2} \vec M_A \cdot \vec M_B$, where $J>0$ describes the AFM coupling between the sublattices.  The nutation amplitude in linear response to the oscillating field $\vec H_{\mathrm{osc}} = \vec h e^{-i\omega t} + \vec h^* e^{i\omega t}$ for weak antiferromagnetic coupling, $J \ll 2\frac{M_0}{\gamma\eta}$,  is then given by 
 \begin{align}
        c_A =&-M_0(\ephi-i\etheta)\ \cdot \bigg(\frac{\vec he^{-i\omega t}}{i(\omega_{\mathrm{n},A}-\omega)+\frac{\alpha}{\eta}} \nonumber \\ &+ \frac{\vec h^{*}e^{i\omega t}}{i(\omega_{\mathrm{n},A}+\omega)+\frac{\alpha}{\eta}} + \frac{\vec H_{0,A}}{i\omega_{\mathrm{n},A} + \frac{\alpha}{\eta}}\bigg),\label{eq:nutWeakA}\\ 
c_B =&M_0(\ephi+i\etheta)\ \cdot \bigg(\frac{\vec h e^{-i\omega t}}{i(\omega_{\mathrm{n}, B} -\omega)+\frac{\alpha}{\eta}}\nonumber \\ &+\frac{\vec h^{*} e^{i\omega t}}{i(\omega_{\mathrm{n}, B} +\omega)+\frac{\alpha}{\eta}}+ \frac{\vec H_{0,B}}{i\omega_{\mathrm{n},B} + \frac{\alpha}{\eta}}\bigg).\label{eq:nutWeakB}
\end{align}
with $\omega_{\mathrm{n},A/B} =  \sqrt{1 + 2J\eta\gamma/M_0}/\eta \pm \gamma \er \cdot \vec H_{\textrm{ext}}$; see Appendix~\ref{sec:nutationAFM} for the derivation. In Eqs.~\eqref{eq:nutWeakA} and \eqref{eq:nutWeakB} the influence of the antiferromagnetic coupling only enters in the form of a renormalized nutation frequency. The nutation frequency is identical to previous results obtained using linear-response theory~\cite{mondal_nutation_2021}. \\ 

By using Eq.~\eqref{eq:dldt} in conjunction with the solution for the nutation vectors Eq.~\eqref{eq:nutWeakA} and \eqref{eq:nutWeakB}, we can derive a differential equation describing the order parameter $\vec N$. The derivation can be found in Appendix~\ref{sec:orderparameterAFM}. The second-order differential equation for the order parameter is 
\begin{align}
\begin{split}
\vec N \times \ddot{\vec N}=&-\gamma\Big[2(\vec N\cdot \vec H_{M})\dot{\vec N}-\vec N\times (\dot{\vec H}_{M}\times \vec N)\Big] \\&- \gamma^2 (\vec N \cdot \vec H_{M})(\vec N \times \vec H_{M}) + \frac{\gamma^2 J}{4}\vec N \times \vec H_{N}\\&+\frac{\mathrm{d}}{\mathrm{d}t}\Big(\vec N \times \vec T_{N}\Big)- \frac{\gamma J}{4}\vec T_M + \gamma (\vec N \times \vec T_{N}) \times \vec H_{M}, \label{eq:d2Ndt2}
\end{split}
\end{align}
where $\vec T_M = \gamma(\partial_t + \frac{\alpha}{\eta})(\Delta \vec L_A + \Delta \vec L_B)$ and $\vec T_{N} =\gamma(\partial_t + \frac{\alpha}{\eta})\Delta\vec{L}_{N}$. $\vec H_M = (\vec H_A + \vec H_B)/2$ contains for example the Zeeman field, whereas $\vec H_N = (\vec H_A - \vec H_B)/2$ takes the anisotropy into account. $\vec H_A$ and $\vec H_B$ are the effective fields acting on the sublattices excluding the exchange interaction.

The last three terms in Eq.~\eqref{eq:d2Ndt2} describe the influence of a finite nutation amplitude $\Delta \vec L_{N}$ on the dynamics. According to Eqs.~\eqref{eq:nutWeakA} and \eqref{eq:nutWeakB}, the total torque exerted by the nutation $\vec T_{N} = \vec T_{N,0} + \vec T_{N,\mathrm{osc}}(t)$ consists of a time-independent part $\vec T_{N,0}$ driven by the Zeeman field and the anisotropy, and an oscillating time-dependent part $\vec T_{N,\mathrm{osc}}(t)$ driven by the oscillating external field. For the first two terms we find
\begin{align}
    \left\langle \frac{\mathrm{d}}{\mathrm{d}t}\Big(\vec N \times \vec T_{N}\Big)- \frac{J}{2}\vec T_M \right \rangle_t = \left\langle \frac{\mathrm{d}}{\mathrm{d}t}\Big(\vec N \times \vec T_{N,0}\Big)- \frac{\gamma J}{4}\vec T_{M,0} \right \rangle_t 
\end{align}
The quickly changing contribution $\vec T_{N,\mathrm{osc}}(t)$ vanishes under time averaging.
Therefore, this term describes damping effects in constant magnetic fields or under the influence of anisotropy, and it is irrelevant for nutational switching, similarly to the term $\langle \frac{\alpha}{\eta} \Delta \vec L \rangle_t$ in Eq.~\eqref{eq:dldtaverage}. Nutational switching is only described by the last term in Eq.~\eqref{eq:d2Ndt2}, which is analogous to $\langle \gamma \Delta \vec L \times \vec  H_{\mathrm{eff}} \rangle_t$ in Eq.~\eqref{eq:dldtaverage}. For further analyzing the impact of nutation on the order parameter, we will neglect all other terms. The second derivative $\ddot{\vec N}$ is then given by
\begin{align}
    \ddot{\vec N} \cdot \etheta = -M_0\bigg(\etheta \cdot \frac{\omega(\omega_{\mathrm n} - \omega) \vec H_{\mathrm{osc}}+\frac \alpha \eta\dot{\vec H}_{\mathrm{osc}}}{(\omega_{\mathrm{n}} - \omega)^2 + \frac{\alpha^2}{\eta^2}} \bigg) \er \cdot \vec H_{\mathrm{osc}} \label{eq:d2ndt2theta}, \\
    \ddot{\vec N} \cdot \ephi = -M_0\bigg(\ephi\cdot \frac{\omega(\omega_{\mathrm n} - \omega) \vec H_{\mathrm{osc}}+\frac \alpha \eta\dot{\vec H}_{\mathrm{osc}}}{(\omega_{\mathrm{n}} - \omega)^2 + \frac{\alpha^2}{\eta^2}} \bigg) \er \cdot \vec H_{\mathrm{osc}}, \label{eq:d2ndt2phi}
\end{align}
see Appendix~\ref{sec:nutationTerm} for the derivation. Here we neglected the influence of the Zeeman field on the nutation frequency, meaning $\omega_{\mathrm{n}, A} = \omega_{\mathrm{n}, B} = \omega_{\mathrm{n}}$. Similarly to FMs, Eqs.~\eqref{eq:d2ndt2theta} and \eqref{eq:d2ndt2phi} show that if the order parameter is perpendicular to the external field, the orientation is stable or unstable. We will further analyze Eqs.~\eqref{eq:d2ndt2theta} and \eqref{eq:d2ndt2phi} for linearly and circularly polarized fields. 

\subsection{Nutational switching for linearly polarized fields}

\begin{figure}
    \centering
    \includegraphics[width = \columnwidth]{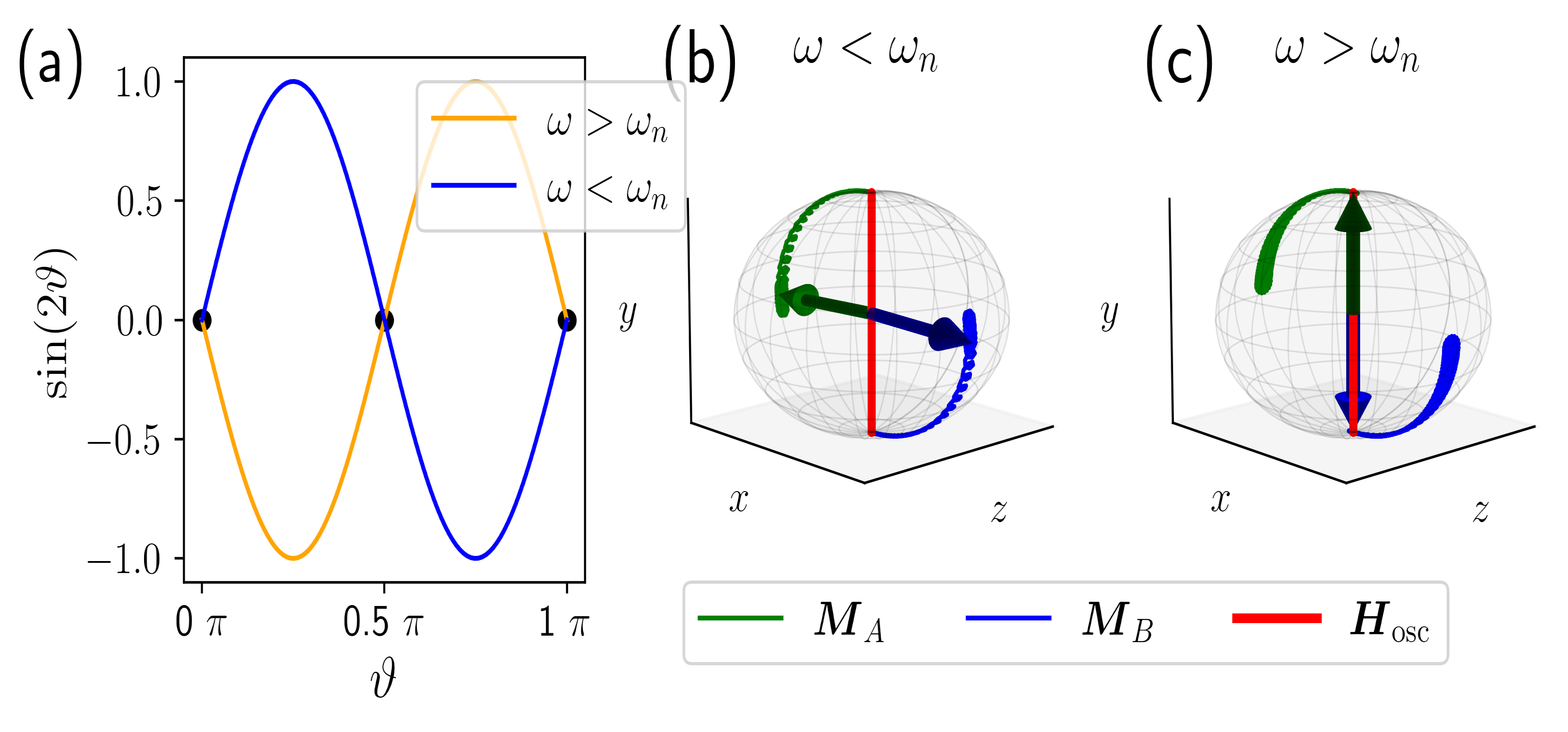}
    \caption{(a) The right-hand side of Eq.~\eqref{eq:afmTilting} as a function of $\vartheta$. (b) and (c) Illustrations of the switching modes when applying a linearly polarized magnetic field to AFMs for external field frequencies $\omega$ (b) below and (c) above the nutation frequency $\omega_{\mathrm{n}}$.
}
    \label{fig:afmLinearModes}
\end{figure}

Next, we investigate the behavior of the order parameter $\vec N$ in a linearly polarized 
magnetic field along the $z$-axis, $\vec H_{\mathrm{osc}} =h \cos(\omega t) \unitvec e_z$. For this purpose, we substitute the magnetic field in Eqs.~\eqref{eq:d2ndt2theta} and \eqref{eq:d2ndt2phi}. For the $\etheta$ component of $\ddot{\vec N}$ we find
\begin{align}
    \ddot{\vec N}\cdot\etheta= \frac{\gamma^2 h^2} 8 \frac{\omega (\omega_{\mathrm{n}} - \omega)}{(\omega_{\mathrm{n}}- \omega)^2 + \frac{\alpha^2}{\eta^2}}\sin(2\vartheta). \label{eq:afmTilting}
\end{align}

Figure~\ref{fig:afmLinearModes}(a) shows an illustration of the right-hand side of Eq.~\eqref{eq:afmTilting}.
If the external frequency $\omega$ is smaller than the nutation frequency $\omega_{\textrm{n}}$, the order parameter is stable if it is perpendicular to the external field at $\vartheta = \pi/2$. $\vec N$ is unstable if it is aligned with the external field at $\vartheta = 0$ or $\vartheta = \pi$. Consequently, the order parameter aligns itself perpendicular to the external field. The qualitative behavior is the same as for FMs. Figure~\ref{fig:afmLinearModes}(b) shows an illustration of this switching mode. For $\omega > \omega_{\mathrm{n}}$ the parallel orientation $\vartheta = 0, \pi$ is stable and the perpendicular orientation $\vartheta =  \pi/2$ is unstable. Therefore, the order parameter aligns itself along the external field. 
Figure~\ref{fig:afmLinearModes}(c) shows an illustration of the latter switching mode.
Differently to FMs, there are two different switching modes depending on the frequency of the external field $\vec H_{\mathrm{osc}}$. These switching modes are also shown in Supplemental Video 3~\cite{supp}.
 
For a linearly polarized field along the $z$-direction, from Eq.~\eqref{eq:d2ndt2phi} immediately follows $\ddot{\vec N} \cdot \unitvec e_\varphi = 0$. Therefore, nutation for linearly polarized fields only leads to tilting but not to a rotation around the $z$-direction.  

Compared to the ferromagnetic case, there is also a different dependence on the material parameters. For $\alpha \to 0$, we get $\frac{\omega_{\mathrm{n}}- \omega}{(\omega_{\mathrm{n}} - \omega)^2 + \frac{\alpha^2}{\eta^2} } \to (\omega_{\mathrm{n}}- \omega)^{-1}$. This is different from the ferromagnetic case, where $\dot \vartheta = 0$ for $\alpha \to 0$. However, in AFMs the tilting vanishes exactly at the resonance frequency and the tilting velocity is the highest for $\omega = \omega_{\mathrm{n}} \pm \alpha/\eta$. 

Next, we use atomistic spin simulations to confirm the analytical prediction that $90\degree$ switching with linearly polarized fields is possible. This switching mode is especially important in AFMs as it can be detected by, e.g., anisotropic magnetoresistance, while a $180\degree$ reorientation of the order parameter cannot be observed experimentally. We consider a system consisting of $4\times 4\times 4$ antiferromagnetically coupled spins in a simple cubic arrangement with free boundary conditions. Furthermore, there is cubic anisotropy described by the Hamiltonian $\mathcal H_{\textrm{ani}} = -K_{4}(M_{Ax}^4 + M_{Bx}^4 + M_{Ay}^4 + M_{By}^4 + M_{Az}^4 + M_{Bz}^4)/M_0^4$. At $t = 0$~ps, the order parameter points along the $z$-direction. We apply a linearly polarized field below the nutation frequency $\omega_{\textrm{n}}$ along the $z$-direction.

\begin{figure}
    \centering
    \includegraphics[width = \columnwidth]{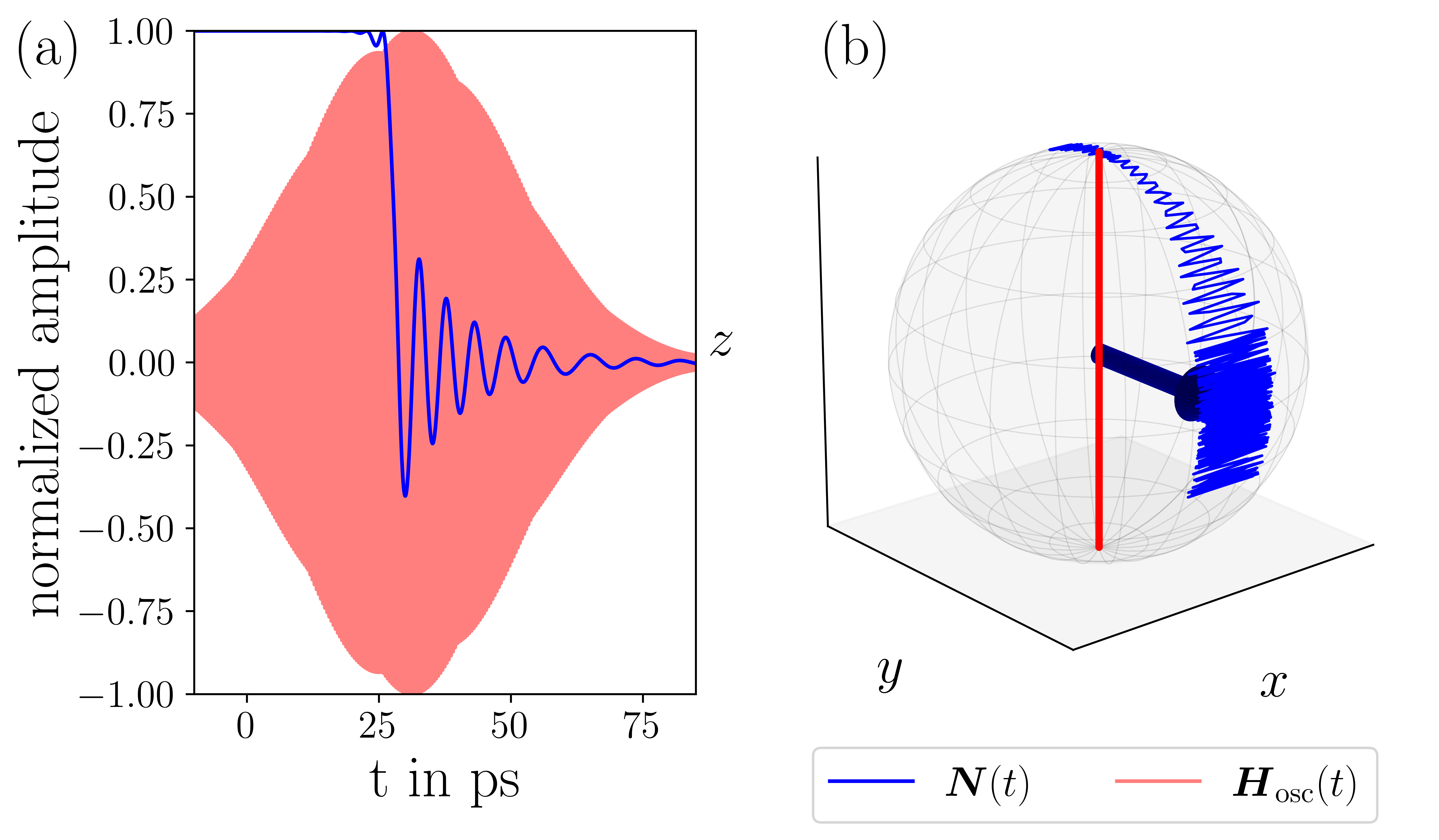}
    \caption{90$\degree$ switching of the order parameter $\vec N$ using a linearly polarized field along the $z$-axis.
The material parameters are $\alpha=0.01, \eta =10$~fs, and a cubic anisotropy with $K_{4}=10^{-24}\ $J. The magnetic field strength is $h=5$~T, the field frequency is $\omega=1\cdot 10^{14}$~s$^{-1}$ and the pulse width is $\text{FWHM} = 47.1$~ps. 
}
    \label{fig:afmLinear}
\end{figure}

Figure~\ref{fig:afmLinear} shows that $90\degree$ switching in AFMs is possible using a linearly polarized field. The pulse width $\text{FWHM} = 47.1$~ps in this simulation is close to the value $47$~ps used in the FM case. However, the switching time, i.e. the time required for the $z$ component of the order parameter to reach $1/\textrm{e}$ of its initial value, is only 5~ps instead of 23~ps. Similarly to FMs, it takes a long time until the nutation is excited as the order parameter is close to an equilibrium orientation. Applying an oscillating field slightly tilted from the equilibrium direction might significantly increase the speed of nutational switching. 

\begin{figure}
    \centering
    \includegraphics[width = \columnwidth]{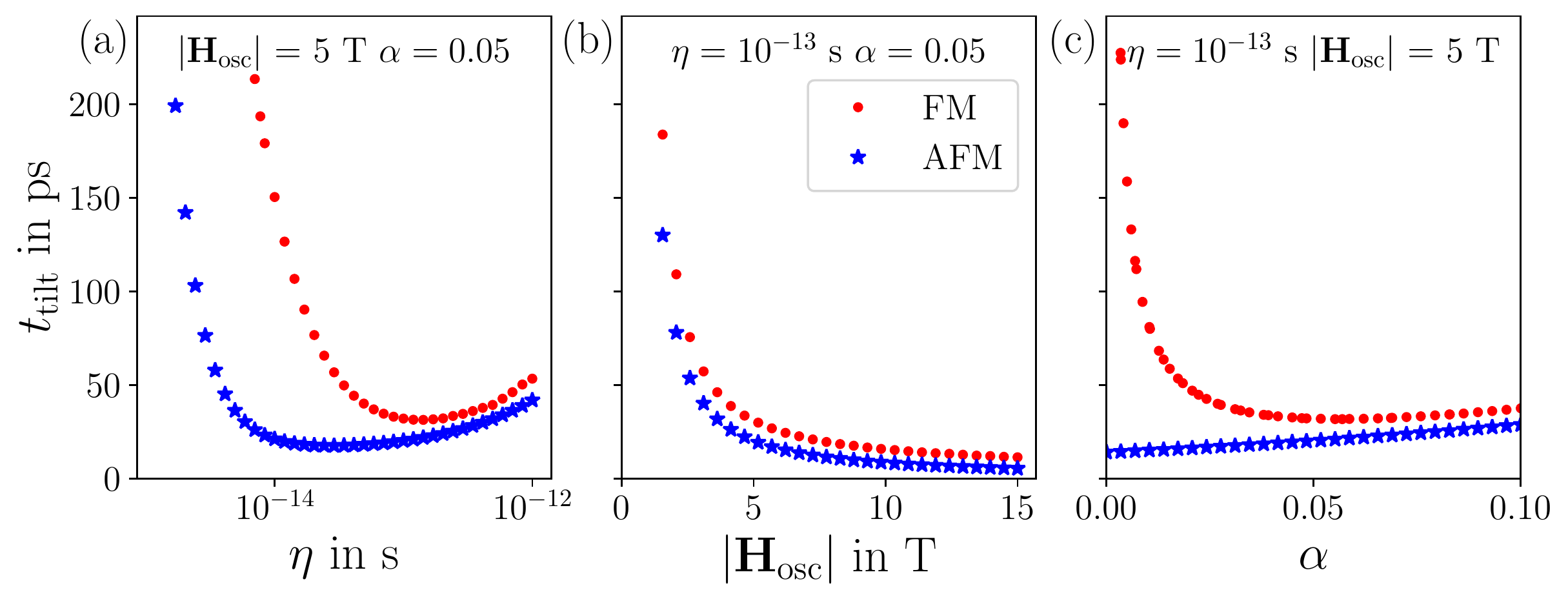}
    \caption{Comparison of the switching times $t_{\mathrm{switch}}$ in FMs and AFMs. Shown is the time needed for the magnetization $M_z$ 
or staggered magnetization $N_z$ to decay to $1/e$ of its initial value. We consider a system without anisotropy or constant external field. The exchange constant is $J_{\mathrm{FM}} = -J_{\mathrm{AFM}} = 1.602 \cdot 10^{-22}$~J. For the FM the field frequency is set to $\omega = 1/\eta$ while for the AFM it is set to $\omega = \sqrt{1 + 2 \gamma J \eta/M_0}/\eta - \frac \alpha \eta$.
}
    \label{fig:tiltComparison}
\end{figure}

Figure~\ref{fig:tiltComparison} shows that the switching time is significantly reduced in AFMs compared to FMs for all values of inertial parameter $\eta$, external field strength $h$, and damping $\alpha$ examined. $90\degree$ switching in AFMs is possible within 10~ps, while in FMs the switching time cannot be lower than 30~ps unless very strong fields are applied. Figure~\ref{fig:tiltComparison}(a) shows that this difference is especially apparent for low values of $\eta$, since the switching time diverges for low inertial parameters. For AFMs, $90\degree$ switching is possible in around 10~ps for $\eta$ between 10~fs and 100~fs. In contrast, in FMs 90$\degree$ switching can only be observed in under 30~ps for $\eta$ around several hundred fs.
Figure~\ref{fig:tiltComparison}(b) shows that nutational switching within a few tens of ps requires field strengths of several tesla in both AFMs and FMs. The lower switching time in AFMs compared to FMs is more pronounced for small field strengths. 
Low values of $\alpha$ lead to diverging switching times in FMs but not in AFMs, see Figure~\ref{fig:tiltComparison}(c) . For FMs, this is not surprising because the tilting velocity $\dot \vartheta$ is proportional to $\alpha$ for resonance according to Eq.~\eqref{eq:tilting}. Therefore, observing nutational switching might be more viable in AFM materials.

\subsection{Circularly polarized fields}

\begin{figure}
    \centering
    \includegraphics[width = \columnwidth]{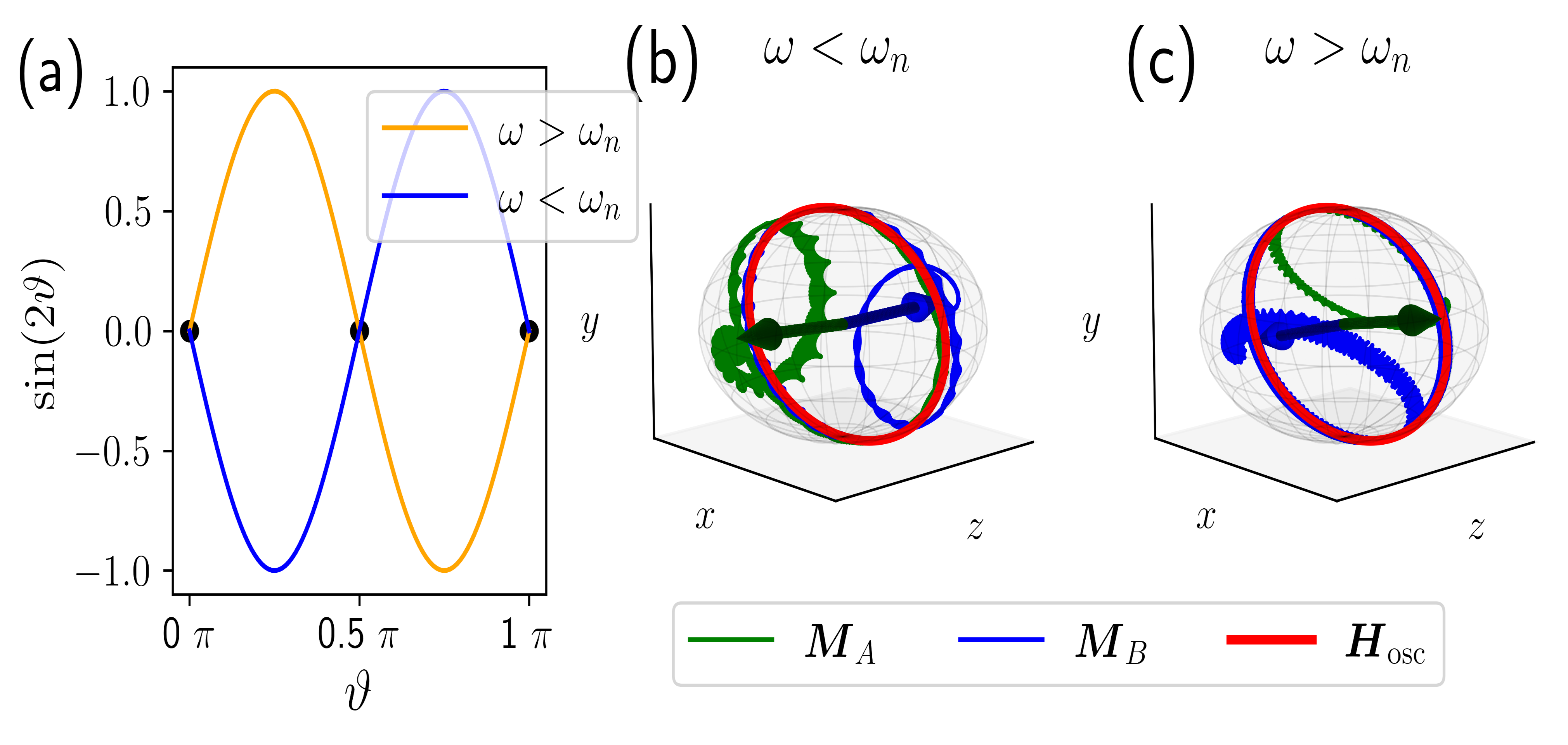}
    \caption{(a) The right-hand side of Eq.~\eqref{eq:afmCircularTheta} as a function of $\vartheta$. (b) and (c) Illustration of the switching modes when applying a circularly polarized magnetic field to AFMs for frequencies $\omega$ (b) below and (c) above the nutation frequency $\omega_{\mathrm{n}}$.
}
    \label{fig:afmAnalyticCircular}
\end{figure}
Next, we investigate the response of the order parameter $\vec N$ to a circularly polarized magnetic field applied in the $xy$-plane. We substitute the magnetic field $\vec H_{\mathrm{osc}} = h \left[ \cos(\omega t)\ \unitvec e_x +  \sin(\omega t)\ \unitvec e_y \right]$ in Eqs.~\eqref{eq:d2ndt2theta} and \eqref{eq:d2ndt2phi} and take the time average over one nutation period. This results in 
\begin{align}
    \ddot{\vec N} \cdot\etheta &= -\frac{\gamma^2 h^2}{4} \, \frac{\left(\omega_{\mathrm{n}} - \omega \right)\omega}{(\omega_{\mathrm{n}} - \omega)^2+\frac{\alpha^2}{\eta^2}} \sin\left(2\vartheta\right), \label{eq:afmCircularTheta} \\\ddot{\vec N}\cdot\ephi &= \gamma^2 h^2 \frac{\frac \alpha \eta \omega}{(\omega_{\mathrm{n}}-\omega)^2 + \frac{\alpha^2}{\eta^2}} \sin(\vartheta). \label{eq:afmCircularPhi}
\end{align}

An illustration of the right-hand side of Eq.~\eqref{eq:afmCircularTheta} is shown in Fig.~\ref{fig:afmAnalyticCircular}(a).
Again different from the FM case, the sign of $\ddot{\vec N} \cdot\etheta$ in Eq.~\eqref{eq:afmCircularTheta} depends on the external field frequency $\omega$.  For a frequency lower than the nutation frequency $\omega < \omega_{\mathrm{n}}$, the order parameter is stable if it is perpendicular to the external field at $\vartheta\ =\ 0$ or $\vartheta = \pi$, and unstable if it is in the plane of the field at $\vartheta = \pi/2$. For all other orientations, the order parameter will align itself perpendicular to the external field, similarly to the FM case. This switching mode is shown in Fig.~\ref{fig:afmAnalyticCircular}(b). For frequencies higher than the nutation frequency $\omega > \omega_{\mathrm{n}}$, the order parameter is stable if it is in the plane of the external field and unstable if it is perpendicular to the magnetic field. Therefore, the order parameter aligns itself in the plane of the external field, where it rotates according to Eq.~\eqref{eq:afmCircularPhi}. This can be used for 180$\degree$ switching, as illustrated in Fig.~\ref{fig:afmAnalyticCircular}(c).

Again in contrast to FMs, in AFMs we find two possible switching modes: from the plane of the excitation towards the perpendicular direction, or the other way around, depending on the excitation frequency. These switching modes are also visualized in Supplemental Video 4~\cite{supp}. This is a consequence of the AFM coupling. In a FM, a state can only be stationary if the torque exerted by the external field vanishes. In AFMs a state can also be stationary if the torques exerted on the two sublattices cancel each other. This is the case for $\vartheta = \pi/2$. Another difference from the FM case is that $\vartheta = 0$ and $\pi$ always have the same stability, instead of one configuration being favored depending on the sense of rotation of the external field. This can be explained by the sublattice symmetry of AFMs. Since the sublattices are identical, the order parameter has to behave the same way if its direction is inverted.
Equation~\eqref{eq:afmCircularPhi} shows that the order parameter rotates around the $z$-axis, similarly to the angular momentum in the single-spin case in Eq.~\eqref{de:PhiCircular}. The rotation vanishes if the order parameter is parallel or antiparallel to the external field.

We also examine the behavior of the order parameter using atomistic spin simulations. Analogously to the FM case, we use circularly polarized pulses for 180$\degree$ switching. We consider a system with uniaxial anisotropy along the $y$-axis, $\mathcal H_{\textrm{ani}} = -\frac{K_y}{M_0^2}\left(M_{Ay}^2 +  M_{By}^2\right)$. At $t = 0$~s, the order parameter is parallel to the $y$-direction. 

\begin{figure}
    \centering
    \includegraphics[width = \columnwidth]{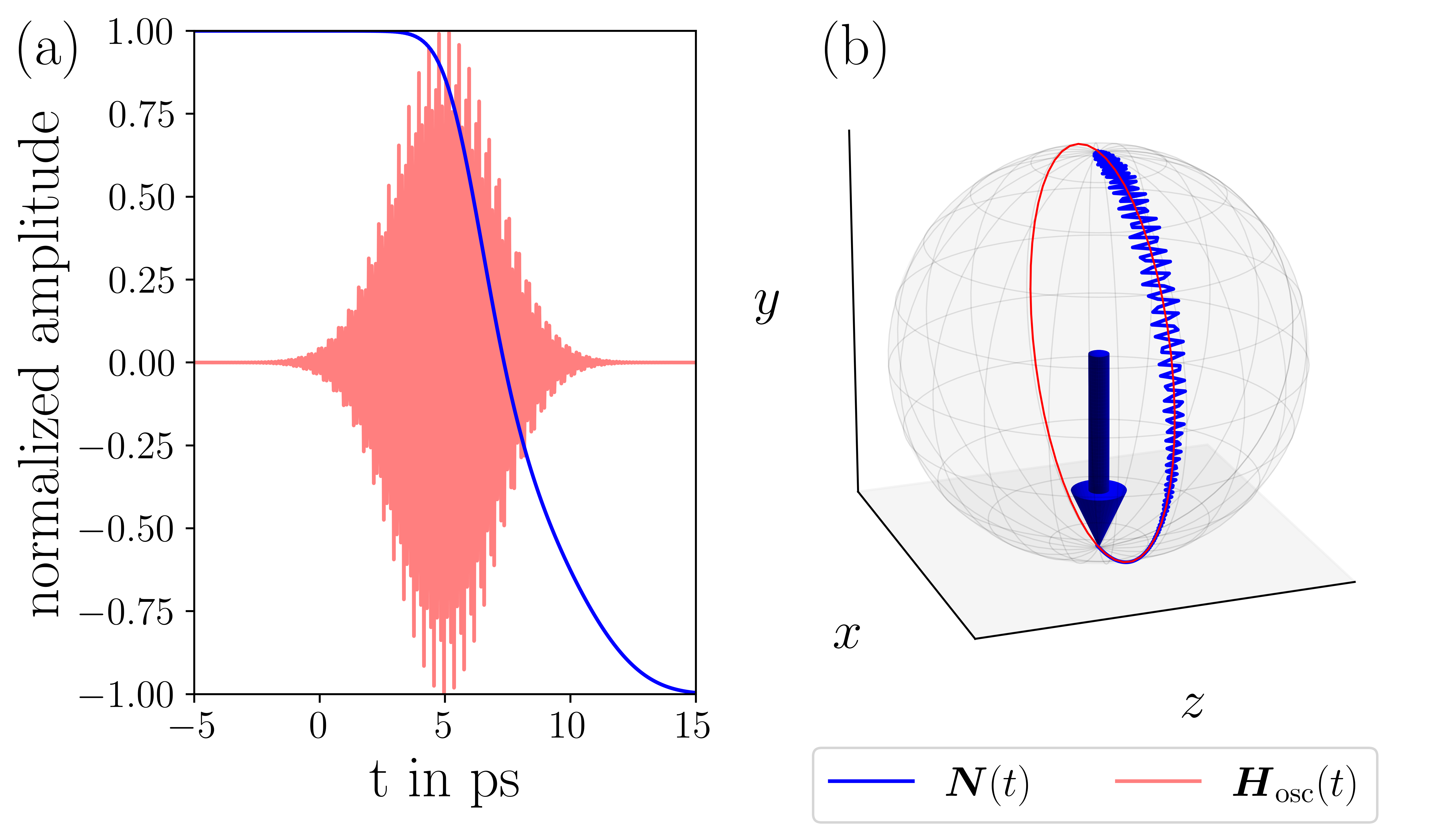}
    \caption{180$\degree$ switching of the order parameter $\vec N$ using a circularly polarized field in the $xy$ plane. The material parameters are $\alpha=0.05, \eta = 10$~fs, $J = 1.602 \cdot 10^{-22}$~J and a uniaxial anisotropy along the $y$-axis with $K_{y} = 10^{-24}$~J. The magnetic field strength is $h=5$~T, the field frequency is $\omega=1.1\cdot 10^{14}$~s$^{-1}$ and the pulse width is $\text{FWHM} = 4.75$~ps. 
}
    \label{fig:afmCircular}
\end{figure}

Figure~\ref{fig:afmCircular} shows that 180$\degree$ switching is possible using circularly polarized fields. Compared to FMs, significantly lower pulse widths are sufficient, $\text{FWHM} = 4.75$~ps instead of $\text{FWHM} = 23.75$~ps in Fig.~\ref{fig:ferroCircularSwitch}. Moreover, the switching itself is possible within 5~ps. This is a significant advantage as the switching process is faster.  Another advantage is that in this case the order parameter is stable in the plane of the magnetic field. Therefore, the switching mode can fully invert the spin without relying on relaxation. Nevertheless, the anisotropy energy is an order of magnitude lower than in the FM example in Fig.~\ref{fig:ferroCircularSwitch}. Atomistic spin simulations indicate that nutational switching in AFMs is only possible for very weak anisotropy.

For a continuously applied magnetic field, the order parameter continuously rotates in the $xy$ plane with the frequency $f_{\varphi}=\dot{\varphi}/(2\pi)$.
To investigate how this frequency varies for different material parameters, we will neglect the uniaxial anisotropy, since otherwise the rotation frequency is not a constant, but varies with the angle $\vartheta$. 
\begin{figure}
    \centering
    \includegraphics[width = \columnwidth]{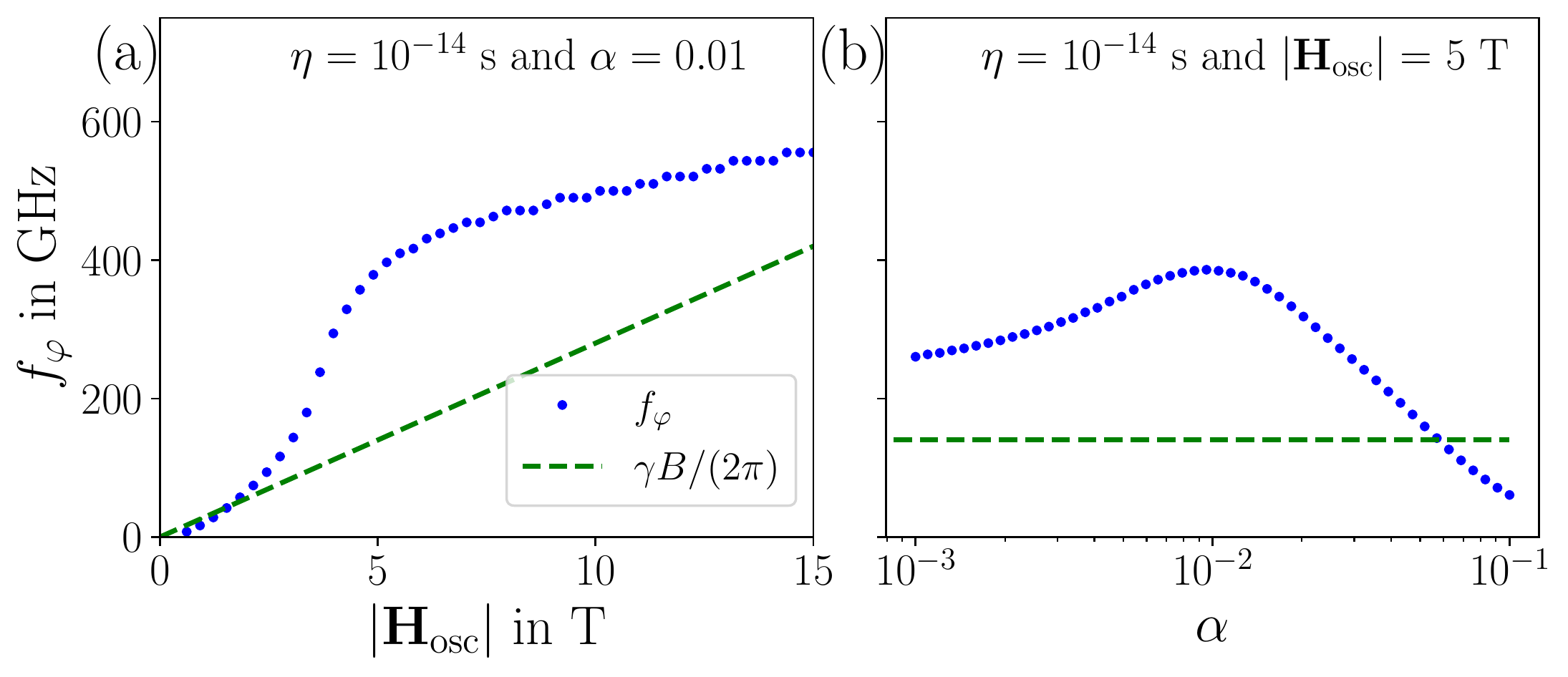}
    \caption{Rotation frequency $f_\varphi$ around the $z$-axis for the first half period. The material parameters are as in Fig.~\ref{fig:afmCircular} but with zero anisotropy.  
A circularly polarized field is continuously applied in the $xy$-plane with $\omega = 1.12 \cdot 10^{14}$ s$^{-1}$. The blue dots show the simulation results. The green dashed line shows the precession frequency for a constant field. 
}
    \label{fig:freqNutationPrecessionComparison}
\end{figure}
Figure~\ref{fig:freqNutationPrecessionComparison}(a) shows that the switching frequency increases with increasing magnetic field, as Eq.~\eqref{eq:afmCircularPhi} implies. Note that Eq.~$\eqref{eq:afmCircularPhi}$ does not give the rotation frequency directly. Therefore, we can expect only qualitative agreement and not quantitatively the same relationship. For field strengths of several tesla, the switching frequency $f_{\varphi}$ significantly exceeds the precession frequency $\omega_{\textrm{p}}=\gamma h$ at comparable magnetic fields. The reason for this is that the resonant excitation of the nutation induces a torque proportional to $h^{2}$, while the velocity of precessional switching is proportional to $h$. A field strength of 5~T switches the spin in approximately 2.5~ps. Using higher field strengths than 5~T provides only diminishing returns, since deviations from the linear-response relation $f_{\varphi}\propto h^{2}$ may be observed in this regime. 

Figure~\ref{fig:freqNutationPrecessionComparison}(b) shows that the switching speed is highest for $\alpha = 0.01$. For very low values of $\alpha$ the switching frequency is diminished but is still significantly higher than for precessional switching. For a value of $\alpha$ greater than 0.01, the switching frequency drops significantly as the nutation becomes suppressed. Therefore, nutational switching can most likely be observed in the low-damping regime. 

\section{Conclusion}

Using analytic methods and atomistic spin simulations of the ILLG equation, we explored nutation-driven switching using oscillating magnetic fields. Analytically we demonstrated that a sinusoidal magnetic field excites a nutational motion of the magnetic moment around the angular momentum, with a resonant enhancement at the nutation frequency. The nutation in conjunction with the oscillating field exerts a torque on the angular momentum. 

This torque can drive either 90$\degree$ or 180$\degree$ nutational switching, depending on the polarization and the frequency of the field. A linearly polarized magnetic field applied to a FM aligns the angular momentum perpendicular to the field. Using atomistic spin simulations, we demonstrated that this enables 90$\degree$ switching using Gaussian pulses for materials with cubic anisotropy. A circularly polarized field in the $xy$-plane aligns the angular momentum along the $z$-direction. The angular momentum continuously rotates in the $xy$-plane, enabling 180$\degree$ switching for Gaussian pulses.

Nutational switching in AFMs is overall similar to FMs for external field frequencies below the nutation frequency, but it proceeds faster because of the intrinsic inertia caused by the exchange interaction between the sublattices. However, for frequencies above the nutation frequency, the order parameter aligns itself parallel to the external field. For circularly polarized fields, this means that the order parameter rotates in the plane of the external field with high frequency. The rotation frequency $\dot\varphi \propto h^{2}$ increases faster than the precession frequency $\omega_{\mathrm{p}} = \gamma h$, enabling faster switching than using precession-based methods. Nutational switching in AFMs is favored in materials with low damping and low anisotropy.

In the future, nutation might be used for faster and more energy-efficient switching of the magnetic state, as the energy stored in the nutation can be extracted during the switching.

\begin{acknowledgments}

The authors would like to thank Ritwik Mondal and Mikhail Cherkasskii for fruitful discussions. Financial support by the German Research Foundation via SFB 1432 and by the National Research, Development and Innovation Office of Hungary via Project No. K131938 is gratefully acknowledged.

\end{acknowledgments}

\appendix

\section{Representing the ILLG equation using the angular momentum} \label{subsec:ILLGAngular}

In this section, we express the ILLG equation~\eqref{eq:ILLG} using the angular momentum $\vec L_{i}$.
The angular momentum is given by~\cite{ciornei_magnetization_2011}
\begin{align}
    \vec L_{i} = \frac{1}{\gamma_{i}} \vec M_{i} -  \frac{ \eta_{i}}{M_{0,i} \gamma_{i}} \vec M_{i} \times \vec{\dot{M}}_{i} = \frac{1}{\gamma_{i}} \vec M_{i} - \Delta \vec L_{i},\label{eq:dl0}
\end{align}
 if an inertial term is considered.

The time derivative of the angular momentum is then described as 
\begin{align}
    \vec{\dot{L}}_{i} = \frac{1}{\gamma_{i}}  \vec{\dot{M}}_{i} -  \frac{ \eta_{i}}{M_{0,i}\gamma} \vec M_{i} \times \vec{\ddot{M}}_{i}. \label{eq:dl1}
\end{align}

Next we eliminate $\vec{\ddot{M}}_{i}$ from Eq.~\eqref{eq:dl1} using the ILLG equation~\eqref{eq:ILLG}. This results in 
\begin{align}
    \gamma_{i} \vec{\dot{L}}_{i} &= \gamma_{i} \left(\frac{1}{\gamma_{i}}\vec{\dot{M}}_{i}  - \frac{\eta}{ M_{0,i}\gamma} \vec M_{i} \times \vec{\ddot{M}}_{i}\right)\nonumber\\&= -\gamma_{i} \vec M_{i} \times \vec H_{\mathrm{eff},i}+ \frac{\alpha_{i}}{M_{0,i}}\vec M_{i} \times \vec{\dot{M}}_{i}. \label{eq:dl2}
\end{align}
Equation~\eqref{eq:dl2} describes that the time evolution of the angular momentum is driven by a precessional torque and a damping term.

By using the fact that $\vec M_{i} \times \vec{\dot{M}}_{i} = \frac{\gamma_{i} M_{0,i}}{ \eta} \parent{\frac{1}{\gamma_{i}} \vec M_{i} - \vec L_{i}}$ according to Eq.~\eqref{eq:dl0}, we obtain
\begin{align}
    \vec{\dot{L}}_{i} = - \vec M_{i} \times \vec H_{\mathrm{eff},i} +  \frac{\alpha_{i}}{\gamma_{i}\eta_{i}} \vec M_{i} - \frac{\alpha_{i}}{\eta_{i}} \vec L_{i}.
\end{align}

Therefore, we found a first-order explicit differential equation for $\vec L_{i}$.
Next, we express $\vec{\dot{M}}_{i}$ from Eq.~\eqref{eq:dl0}. This can be done by multiplying the equation by $\vec M_{i} \times$ and using the triple vector product
\begin{align}
    \vec M_{i} \times \vec L_{i} =&  -\frac{\eta_{i}}{\gamma_{i} M_{0,i}} \parent{\underbrace{\parent{\vec M_{i} \cdot \vec{\dot{M}}_{i}}}_{ = 0} \vec M_{i} \nonumber\\& - \underbrace{\parent{\vec M_{i} \cdot \vec M_{i}}}_{= M_{0,i}^2} \vec{\dot{M}}_{i}}  = \frac{\eta_{i} M_{0,i}}{\gamma_{i}} \vec{\dot{M}}_{i}.
\end{align}
Here, we used that the ILLG equation conserves the length of the magnetic moment, meaning that $\vec M_{i}$ and $\vec{\dot{M}}_{i}$ are perpendicular to each other.

The new equations of motion can be written as
\begin{align}
    \vec{\dot{L}}_{i} &= - \vec M_{i} \times \vec H_{\mathrm{eff},i} + \frac{\alpha_{i}}{ \eta_{i}} \left( \frac{1}{\gamma_{i}} \vec M_{i}- \vec L_{i} \right),  \label{eq:dldt2}\\
    \vec{\dot{M}}_{i} &= \frac{\gamma_{i}}{\eta_{i} M_{0,i}} \vec M_{i} \times \vec L_{i}. \label{eq:dmdt}
\end{align}

According to Eq.~\eqref{eq:dl0}, we can write $\vec M_{i}$ as $\vec M_{i} = \gamma_{i} (\vec L_{i} + \Delta \vec L_{i})$. By substituting this into Eq.~\eqref{eq:dmdt} we obtain
\begin{align}
    \vec{\dot{L}}_{i} + \Delta\vec{\dot{L}}_{i} = \frac{\gamma_{i}}{\eta_{i} M_{0,i}} \Delta \vec L_{i} \times \vec L_{i}.
\end{align}

Now we can use Eq.~\eqref{eq:dldt2} to eliminate $\vec{\dot{L}}_{i}$, resulting in Eqs.~\eqref{eq:dldt} and \eqref{eq:ddeltaLdt} in the main text.

\section{Differential equation for the nutation vector} \label{sec:difEquationNutation}

We will assume that the system is originally in equilibrium, where $\vec L_{i}$ and $\vec M_{i}$ are parallel. The nutation is excited by the oscillating external field, which is assumed to be small in order to remain in the linear-response regime. While the length of $\vec M_{i}$ is conserved in Eq.~\eqref{eq:ILLG}, in the limit of low nutation amplitude it can also be assumed that the magnitude of $\vec{L}_{i}$ is conserved on the time scale of the order of the nutation period. We set $\vec L_{i} = L_{0,i} \unitvec e_{r,i}$, and we assume $\Delta \vec L_{i} = a_{i}(t) \unitvec e_{\vartheta,i} + b_{i}(t) \unitvec e_{\varphi,i}$. The assumption that the nutation is in the plane perpendicular to $\vec L_{i}$ is justified for a small nutation amplitude, because $\vec L_{i} \cdot \Delta \vec L_{i} = -|\Delta \vec L_{i}|^2 \approx 0$.  We substitute the approximation for $\Delta \vec L_{i}$ into Eq.~\eqref{eq:ddeltaLdt}. Furthermore, we assume that the time derivatives of the basis vectors $\unitvec e_{r,i}$, $\unitvec e_{\varphi,i}$, and $\unitvec e_{\vartheta,i}$ can be neglected on the fast time scale on which the nutation takes place. This results in
\begin{align}
\begin{split}
        \dot{a}_{i} 
        \unitvec e_{\vartheta,i} + \dot{b}_{i} 
        \unitvec e_{\varphi,i} 
    =& \frac{M_{0,i}}{\eta_{i} M_{0,i}} \bigg(-a_{i}\unitvec e_{\varphi,i} + b_{i}\unitvec e_{\vartheta,i} \bigg) \\ &+ M_{0,i} \unitvec e_{r,i} \times \vec H_{\mathrm{eff},i} + \gamma_{i} \bigg(a_{i} \unitvec e_{\vartheta,i} \times \vec H_{\mathrm{eff},i} \\ &+ b_{i} \unitvec e_{\varphi,i} \times \vec H_{\mathrm{eff},i}\bigg) - \frac{\alpha_{i}}{\eta_{i}} (a_{i}\unitvec e_{\vartheta,i} + b_{i} \unitvec e_{\varphi,i}).
\end{split}
\end{align}

 By projecting on the $\unitvec e_{\vartheta,i}$ and $\unitvec e_{\varphi,i}$ directions we obtain 
\begin{align}
    \dot{a}_{i} 
    &= -\frac{\alpha_{i}}{\eta_{i}} a_{i} + \frac{L_{0,i}\gamma_{i}}{M_{0,i} \eta_{i}} b_{i} + \gamma_{i} b_{i} \vec H_{\mathrm{eff},i} \cdot \unitvec e_{r,i} - M_{0,i} \unitvec e_{\varphi,i} \cdot \vec H_{\mathrm{eff},i}, \qquad \label{eq:nutA}\\
    \dot{b}_{i} 
    &= -\frac{\alpha_{i}}{\eta_{i}} b_{i} - \frac{L_{0,i}\gamma_{i}}{M_{0,i} \eta_{i}} a_{i} - \gamma_{i} a_{i} \vec H_{\mathrm{eff},i} \cdot \unitvec e_{r,i} + M_{0,i} \unitvec e_{\vartheta,i} \cdot \vec H_{\mathrm{eff},i}. \qquad\label{eq:nutB} 
\end{align}

To express $L_{0,i}$, we use the approximation
\begin{align}
\begin{split}
    L_{0,i}^2 &= \left| \frac{1}{\gamma_{i}} \vec M_{i} - \frac{\eta_{i}}{M_{0,i}\gamma} \vec M_{i} \times \vec{\dot{M}}_{i} \right|^2 = \frac{1}{\gamma_{i}^2}|\vec M_{i}|^2  + \frac{\eta_{i}^2}{\gamma_{i}^2}\left|\vec{\dot{M}}_{i}\right|^2 \\ &\approx \frac{1}{\gamma_{i}^2}|\vec M_{i}|^2 = \frac{M_{0,i}^2}{\gamma_{i}^2} \end{split}\label{eq:L0}.
\end{align}

Here we used the fact that $\vec M_{i} \cdot (\vec M_{i} \times \vec{\dot{M}}_{i}) = 0$ and $\vec M_{i} \perp \vec{\dot{M}}_{i}$. This approximation becomes exact for vanishing nutation amplitude. Using this we obtain $\frac{L_{0,i} \gamma_{i}}{M_{0,i} \eta_{i}}\approx \frac{1}{\eta_{i}}$. Next we combine Eqs.~\eqref{eq:nutA} and \eqref{eq:nutB} into
\begin{align}
    \partial_{t}(a_{i} + ib_{i})  =& -\frac{\alpha_{i}}{\eta_{i}} (a_{i} + ib_{i}) -\frac{L_{0,i}\gamma_{i}}{M_{0,i}\eta_{i}} (ia_{i} - b_{i})\nonumber\\& -\gamma_{i}(ia_{i} - b_{i}) \vec H_{\mathrm{eff},i} \cdot \unitvec e_{r,i} \nonumber\\&- M_{0,i}(\unitvec e_{\varphi,i}  - i\unitvec e_{\vartheta,i}) \cdot \vec H_{\mathrm{eff},i}.
\end{align}

By defining $c_{i}(t) = a_{i}(t) + ib_{i}(t)$, this simplifies to 

\begin{align}
\begin{split}
    \dot{c}_{i}
    =& - i\omega_{\textrm{n},i} c_i(t) - \frac{\alpha_i}{\eta_i}c_i(t) \\ &- M_{0,i}(\unitvec e_{\varphi, i}  - i\unitvec e_{\vartheta, i}) \cdot \vec H_{\mathrm{eff}, i}, \label{de:nutation}
\end{split}
\end{align}
with $\omega_{\textrm{n},i}=\frac{1}{\eta_{i}}+\gamma_{i}\vec H_{\mathrm{eff}, i} \cdot \er$ the nutation frequency. 

Next, we will solve Eq.~\eqref{de:nutation} for a single macrospin describing a ferromagnetic particle. As there is only one sublattice, we drop the index $i$. Assuming that $\vec H_{\mathrm{eff}}$ does not depend on the nutation amplitude $c(t)$, a closed-form solution can be calculated. The solution to the homogeneous part is given by 
\begin{align}
    c_{\mathrm{hom}}(t) =& c_0 e^{- \frac \alpha \eta t}e^{-i\int_{0}^{t}\omega_{\mathrm{n}}(t')\textrm{d}t'}.\label{eq:homNutation}
\end{align}

The nutation is a circular motion with exponentially decaying amplitude.

In the presence of an effective field $\vec{H}_{\mathrm{eff}}$, the solution to the inhomogeneous equation is given by 
\begin{align}
    c(t)= c_{\mathrm{hom}}(t)\parent{1 - M_0\int_0^t dt'\  c^{-1}_{\mathrm{hom}}(t')\ (\unitvec e_\varphi  - i\unitvec e_\vartheta) \cdot \vec H_{\mathrm{eff}}(t')}. \label{eq:inhomNutation}
\end{align}

In the main text, we {{rewrote Eq.~\eqref{eq:inhomNutation} in the form of Eq.~\eqref{eq:nutationClosedForm}}} for analyzing the nutation amplitude in the presence of anisotropy, a static field, and an oscillating field. 

\section{Connection between damping and nutation} \label{sec:nutationDamping}
For a system with an oscillating field, uniaxial anisotropy, and a static field, we can explicitly evaluate the damping term $\braket{\frac \alpha \eta \Delta \vec L}_t$ in Eq.~\eqref{eq:dldtaverage}. For this purpose, we use the closed-form solution for the nutation Eq.~\eqref{eq:inhomNutation}.
By using $\omega_{\mathrm{n}} \approx \frac{1}{\eta}$ and $\braket{\vec H_{\mathrm{eff}}}_t = \braket{\vec H_{\mathrm{osc}}(t) + \vec H_0}_t = \vec H_0$, we arrive at 
\begin{align}
\begin{split}
    \left\langle  \frac{\alpha}{\eta}\Delta \vec L\right\rangle_t = M_0\frac{\alpha}{1 + \alpha^2}\bigg[(\etheta \cdot \vec H_0)\etheta + (\ephi \cdot \vec H_0)\ephi \\ + \alpha\left[(\etheta \cdot \vec H_0) \ephi - (\ephi \cdot \vec H_0)\etheta\right] \bigg] \label{eq:damping}
\end{split}
\end{align}

The first two terms describe the same damping process as in the LLG equation. The angular momentum will align itself parallel to the magnetic field $\vec H_0$. The speed of the relaxation is controlled by the parameter $\alpha$. Note that Eq.~\ \eqref{eq:damping} does not explicitly depend on $\eta$, demonstrating that there are no inertial effects in this case. The third and fourth terms on the right-hand side of Eq.~\eqref{eq:damping} describe a damping-related modification to the precession frequency. 

\section{Nutation in antiferromagnets} \label{sec:nutationAFM}
We express the angular momenta of the sublattices in spherical coordinates $\vec L_A = L_0 \unitvec e_{r, A}$ and $\vec L_B = L_0 \unitvec e_{r, B}$. In AFMs, the total magnetization is minimized by the AFM coupling, $\vec M = \vec M_A + \vec M_B \approx 0$. Therefore, the angular momenta of the two sublattices must be approximately antiparallel, $\vec L_A \approx - \vec L_B$. This enables us to express the orientation of the sublattices in terms of two global angles $\vartheta = \vartheta_A$ and $\varphi = \varphi_A$. These angles define the orientation of the order parameter $\vec N = \vec M_A - \vec M_B$. For the basis vectors this means $\hat{\boldsymbol{e}}_{r,A} = -\hat{\boldsymbol{e}}_{r,B}, \hat{\boldsymbol{e}}_{\vartheta,A} = \hat{\boldsymbol{e}}_{\vartheta,B}$, and $\hat{\boldsymbol{e}}_{\varphi,A} = -\hat{\boldsymbol{e}}_{\varphi,B}$. Here the perpendicular component of the effective field $(\unitvec e_{\varphi, i}  - i\unitvec e_{\vartheta, i}) \cdot \vec H_{\mathrm{eff}, i}$ in Eq.~\eqref{de:nutation} contains terms which are linear in $c_{A}$ and $c_{B}$ because of the AFM exchange interaction. This leads to the following system of differential equations: 
\begin{align}
    \dot{c}_A &=  - i\omega_{A} c_A - \frac{\alpha}{\eta}c_A  -i\gamma^2 L_0 J c_B^* + f_{A}(t), \label{eq:ca}\\
     \dot{c}^{*}_B &= - i\omega_{B} c_B^{*} - \frac{\alpha}{\eta}c_B^{*} + i\gamma^2 L_0 J c_A + f_B^{*}(t), \label{eq:cb}
\end{align}
with $\omega_{A/B} = \gamma \vec H_{\textrm{ext}} \cdot \er   \pm (\frac{1}{\eta}  - \gamma \vec H_{\textrm{ani},A/B} \cdot \er + \gamma^2 L_0 \frac{J}{M_0^2})$ and $f_{A/B} = M_0(\ephi \mp i\etheta)\cdot (\vec H_{\textrm{ext}}+\vec H_{\textrm{ani},A/B})$.

 The homogeneous solution to this system of differential equations can be obtained by diagonalizing the corresponding dynamical matrix, yielding 
\begin{align}
    \left(\begin{array}{c} c_{A}\\ c^{*}_{B} \end{array}\right) &=  \left(\begin{array}{c}  \sigma_A\\ 1 \end{array}\right) e^{-i\omega_{\textrm{n},A}t}e^{-\frac{\alpha}{\eta}t} + \left(\begin{array}{c} \sigma_B\\ 1 \end{array}\right) e^{i\omega_{\textrm{n},B} t }e^{-\frac{\alpha}{\eta}t} . \label{eq:nutAFMHom}
\end{align}
with $A = (M_0 - 2K(\vec u \cdot \er)^2\eta\gamma)\cdot (M_0 + 2J\eta \gamma - 2K(\vec u \cdot \er)^2\eta \gamma)$, $\omega_{\textrm{n},A/B} = (\frac{\sqrt A}{M_0\eta} \pm \gamma \er \cdot \vec H_0)$, 
$\sigma_B=\frac{M_{0}+J\,\eta \,\gamma -2\,K\,\eta \,\gamma(\er \cdot \vec u)^2 -\sqrt A}{J\,\eta \,\gamma }$ and $\sigma_A\ = \sigma_B + 2\frac{\sqrt A}{J\eta \gamma}$. 

\begin{align}
\begin{split}
        c_A &=M_0\frac{(\ephi-i\etheta)}{\sigma_{A} - \sigma_{B}}\cdot \bigg[\frac{\sigma_{B} \vec h e^{-i\omega t}}{i(\omega_{\mathrm{n}, B}-\omega)+\frac{\alpha}{\eta}} + \frac{\sigma_{A} \sigma_{B} \vec h^* e^{i\omega t}}{i(\omega_{\mathrm{n}, B}+\omega)+\frac{\alpha}{\eta}}\bigg)
\\ &-\frac{\sigma_{A}\vec h e^{-i\omega t}}{-i(\omega_{\mathrm{n}, A}+\omega)+\frac{\alpha}{\eta}} -\frac{\sigma_{A}\sigma_{B} \vec h^* e^{i\omega t}}{-i(\omega_{\mathrm{n}, A}-\omega)+\frac{\alpha}{\eta}}\bigg],\label{eq:nutationAFMSolA} \end{split}\\
\begin{split}
c_B &= M_0\frac{(\ephi+i\etheta)}{\sigma_{A} - \sigma_{B}}\cdot \bigg[\frac{\vec h^* e^{i\omega t}}{-i(\omega_{\mathrm{n}, B}-\omega)+\frac{\alpha}{\eta}} + \frac{\sigma_{A} \vec h e^{-i\omega t}}{-i(\omega_{\mathrm{n}, B}+\omega)+\frac{\alpha}{\eta}}
\\  &- \frac{\vec h^* e^{i\omega t}}{i(\omega_{\mathrm{n}, A}+\omega)+\frac{\alpha}{\eta}}- \frac{\sigma_{B}\ \vec h e^{-i\omega t}}{i(\omega_{\mathrm{n}, A}-\omega)+\frac{\alpha}{\eta}}\bigg].\label{eq:nutationAFMSolB}
\end{split}
\end{align}

Discussing nutational switching based on the above equation is rather convoluted. Therefore, we will only consider the case where $\omega_{\mathrm{n}, A} \approx \omega_{\mathrm{n}, B} \approx \frac{\sqrt A}{M_0 \eta }$ which is usually the case as the nutation frequency $\frac{\sqrt A}{M_0\eta }$ is much greater than the precession frequency $\gamma |\vec H_0|$. Moreover, for $\sigma_{A} \to 0$ the nutation modes of the sublattices Eq.~\ \eqref{eq:nutAFMHom} decouple. Following a Taylor expansion of $\sqrt A$ in $J\eta\gamma$ up to the second order, we find $\sigma_{A} = \frac{J\eta \gamma }{ 2(M_0-2K\eta \gamma (\hat e_r \cdot \vec u)^2)}$. 
If the anisotropy is weak, the sublattices decouple under the condition $J \ll 2\frac{M_0}{\eta \gamma }$. In the simulations we considered $\eta = 10~\mathrm{fs}$ and $M_0 = 1.8548 \cdot 10^{-23}\frac{\mathrm J}{\mathrm T}$, requiring $J \ll 1.055\cdot 10^{-20}$~J. For the value of $J=1.602\cdot 10^{-22}$~J this approximation is well justified. In this case, Eqs.~\eqref{eq:nutationAFMSolA}, \eqref{eq:nutationAFMSolB} have the same form as the single-spin nutation solution Eq.~\eqref{eq:nutationClosedForm}.

\section{Order parameter in AFMs} \label{sec:orderparameterAFM}
In this section, we derive the differential equation~\eqref{eq:d2Ndt2} for the order parameter $\vec N$ in AFMs. The derivation is similar to a previous work by Gomona\u{\i} \textit{et al.}~\cite{gomonai_loktev_1970} without the inertial term. We add $ \Delta \vec{\dot{L}}_{A/B}$ to Eq.~\eqref{eq:dldt} and multiply by $\gamma_{A/B}=\gamma$. This results in the following differential equations for the magnetizations $\vec M_{A/B}$ of the different sublattices:
\begin{align}
\vec{\dot{M}}_A&=-\gamma \vec M_A \times \vec H_{A} + \frac{\gamma J}{M_0^2} \vec M_A \times \vec M_B+ \left(\partial_t+\frac{\alpha}{\eta}\right)\gamma \Delta \vec L_A, \label{eq:MA} \\
\vec{\dot{M}}_B&=-\gamma \vec M_B \times \vec H_{B} + \frac{\gamma J}{M_0^2} \vec M_B \times \vec M_A+ \left(\partial_t+\frac{\alpha}{\eta}\right)\gamma \Delta \vec L_B, \label{eq:MB} 
\end{align}
with $\vec{H}_{A/B}=\vec{H}_{\textrm{ext}}+\vec{H}_{\textrm{ani},A/B}$. Next, we define the magnetization $\vec M = \vec M_A + \vec M_B$ and the staggered magnetization $\vec N = \vec M_A - \vec M_B$.  The effective fields for the magnetization and staggered magnetization are given by $\vec H_{M} = (\vec H_A + \vec H_B)/2$ and $\vec H_{N} = (\vec H_A - \vec H_B)/2$. By summing up and subtracting Eqs.~\eqref{eq:MA} and \eqref{eq:MB}, we find a system of differential equations for $\vec M$ and $\vec N$:
\begin{align}
\vec{\dot{M}}&=-\gamma (\vec M \times \vec H_{M} + \vec N \times \vec H_{N}) + \vec T_M, \label{de:dm} \\ 
\vec{\dot{N}}&=-\gamma (\vec N \times \vec H_{M} + \vec M \times \vec H_{N}) + \frac{\gamma J}{M_0^2}
\vec N \times \vec M + \vec T_{N}, \label{de:dN}
\end{align}
with 
$\vec T_M = (\partial_t + \frac{\alpha}{\eta})\left(\Delta \vec{L}_{A}+\Delta \vec{L}_{B}\right)$ and
$\vec T_{N} = (\partial_t + \frac{\alpha}{\eta})\left(\Delta \vec{L}_{A}-\Delta \vec{L}_{B}\right)$. 

By multiplying Eq.~\eqref{de:dN} with $\vec N \times$ and applying the triple vector product, we can find an explicit expression for $\vec M$. Using the fact that for antiferromagnets $\vec M \cdot \vec N = 0$, and the approximation that the anisotropy is much smaller than the exchange interaction, we find the expression
\begin{align}
\vec M \approx -\frac{4}{\gamma J }
\bigg(\vec N \times \dot{\vec N} + \gamma \vec N\times \left(\vec N\times \vec H_{M}\right)-\vec N \times \vec T_{N}\bigg).
\end{align}

We can now substitute this into Eq.~\eqref{de:dm}. This reveals
\begin{align}
\begin{split}
&\vec N \times \ddot{\vec N}+\gamma \partial_{t}\bigg(\vec N \times \left(\vec N \times \vec H_{M}\right) - \vec N \times \vec T_{N}\bigg) \\&-\frac{\gamma^2 J}{4} 
\vec N \times \vec H_{N}  =-\gamma\bigg(\vec N \times \dot{\vec N} + \gamma \vec N\times \left(\vec N\times \vec H_{M}\right)\\&-\vec N \times \vec T_{N}\bigg)\times \vec H_{M}  - \frac{\gamma J}{4}\vec T_M. \label{eq:d2ndt2prev}
\end{split}
\end{align}

We can now simplify
$\vec N \times \left(\vec N \times \vec H_{M}\right) = (\vec N \cdot \vec H_{M}) \vec N - 4M_0^2 \vec H_{M}$.
Therefore, we obtain
$(\vec N \times (\vec N \times \vec H_{M})  )\times \vec H_{M} = (\vec N \cdot \vec H_{M})(\vec N \times\vec H_{M}).$
Moreover, using the assumption
that the nutation $\Delta\vec{L}_{N}$ occurs in the plane perpendicular to $\vec{N}$, we obtain $\vec{N}\cdot\vec{T}_{N}\approx 0$ and $\vec{N}\cdot\dot{\vec{N}}\approx 0$ from Eq.~\eqref{de:dN}, i.e., the length of the staggered magnetization is conserved. This implies $\vec{N} \times (\dot{\vec{N}} \times \vec H_{M}) \approx \left(\vec{N}\vec{H}_{M}\right)\dot{\vec{N}}$. 
This simplifies Eq.~\eqref{eq:d2ndt2prev} to Eq.~\eqref{eq:d2Ndt2} in the main text.

\section{Nutation term in antiferromagnets in spherical coordinates} \label{sec:nutationTerm}

In antiferromagnets, the term $\gamma (\vec N \times \vec T_{N}) \times \vec H_{\mathrm M}\approx \gamma (\vec N \times \partial_t \Delta \vec L_N)\times \vec H$ in Eq.~\eqref{eq:d2ndt2prev} is responsible for nutational switching. Here, we evaluate this term using the closed-form solution for the nutation vector in Eqs.~\eqref{eq:nutationAFMSolA} and \eqref{eq:nutationAFMSolB}.  We assume $\omega_{\mathrm{n},A} \approx \omega_{\mathrm{n}, B} \approx \omega_{\mathrm{n}}$ which is valid as the precession frequency is much smaller than the nutation frequency. First, we know that $\Delta \vec L_N =\textrm{Re}\{c_A\} \etheta + \textrm{Im}\{c_A\} \ephi - (\textrm{Re}\{c_B\} \etheta - \textrm{Im}\{c_B\} \ephi ) = \textrm{Re}\{c_A - c_B^*\} \etheta + \textrm{Im}\{c_A - c_B^*\} \ephi$. According to Eqs.~\eqref{eq:nutWeakA} and \eqref{eq:nutWeakB},  
\begin{align}
    &c_A-c_B^*=M_0\bigg\{\frac{-\ephi + i\etheta}{(\omega_{\mathrm{n}} - \omega)^2 + \frac{\alpha^2}{\eta^2}}\cdot \bigg[i(\omega_{\mathrm{n}} - \omega)\nonumber\\ 
&(-\vec h e^{-i\omega t}+\vec h^* e^{i\omega t})+\frac \alpha \eta (\vec h e^{-i\omega t}+\vec h^* e^{i\omega t})\bigg]\bigg\} \nonumber\\ 
&=M_0\bigg[\frac{-\ephi + i\etheta}{(\omega_{\mathrm{n}} - \omega)^2 + \frac{\alpha^2}{\eta^2}}\cdot \bigg(\frac{\omega_{\mathrm{n}} - \omega}{\omega} \dot{\vec H}_{\mathrm{osc}}+\frac \alpha \eta \vec H_{\mathrm{osc}}\bigg) \bigg]. \label{eq:c_a-c_b}
\end{align}

$\vec T_{N}$ can be approximated in the following way:
\begin{align}
    \vec T_{N} \approx \partial_t \Delta \vec L_N= -M_0\bigg[\ephi\cdot \frac{(\omega_{\textrm{n}} - \omega)\omega \vec H_{\mathrm{osc}}+\frac{\alpha}{\eta}\dot{\vec H}_{\mathrm{osc}}}{(\omega_{\textrm{n}} - \omega)^2 + \frac{\alpha^2}{\eta^2}} \bigg]\etheta \\\qquad\qquad\qquad+ M_0\bigg[\etheta\cdot \frac{(\omega_{\textrm{n}} - \omega)\omega \vec H_{\mathrm{osc}}+\frac{\alpha}{\eta}\dot{\vec H}_{\mathrm{osc}}}{(\omega_{\textrm{n}} - \omega)^2 + \frac{\alpha^2}{\eta^2}} \bigg]\ephi.
\end{align}

Here we used the fact that $\ddot{\vec H}_{\mathrm{osc}} = -\omega^2 \vec H_{\mathrm{osc}}$, where $\vec H_{\mathrm{osc}} = \vec h e^{-i\omega t} + \vec h^* e^{i\omega t}$ is the time-dependent oscillating field. Using Eq.~\eqref{eq:c_a-c_b} and $\vec N = 2M_0 \cdot \er$, we can write Eq.~\eqref{eq:d2Ndt2} in the limit where only the nutation term is kept on the right-hand side as
\begin{align}
    \vec N \times \ddot{\vec N} \approx & -2M_0\bigg[\ephi\cdot \frac{\omega(\omega_{\mathrm{n}} - \omega) \vec H_{\mathrm{osc}} +\frac \alpha \eta \dot{\vec H}_{\mathrm{osc}}}{(\omega_{\mathrm{n}} - \omega)^2 + \frac{\alpha^2}{\eta^2}} \bigg]( \ephi \times \vec H_{\mathrm{osc}}),\nonumber\\
 &- 2M_0\bigg[\etheta\cdot \frac{\omega(\omega_{\mathrm{n}} - \omega) \vec H_{\mathrm{osc}}+\frac \alpha \eta \dot{\vec H}_{\mathrm{osc}}}{(\omega_{\mathrm{n}} - \omega)^2 + \frac{\alpha^2}{\eta^2}} \bigg](\etheta \times \vec H_{\mathrm{osc}}).
\end{align}

The projections of $\ddot{\vec N}$ on the $\etheta$ and $\ephi$ directions are then given by Eqs.~\eqref{eq:d2ndt2theta} and \eqref{eq:d2ndt2phi}, respectively.

\end{document}